\documentclass{aa}
\usepackage{graphicx}
\usepackage{natbib}
\bibpunct{(}{)}{;}{a}{}{,}
\begin{document}
   \title{Intensity contrast from MHD simulations and from HINODE observations}
  
   \author{N.~Afram\inst{1}
          \and Y.~C.~Unruh\inst{1}
          \and S.~K.~Solanki\inst{2,3}
           \and M.~Sch\"ussler\inst{2}
	   \and A.~Lagg\inst{2}
	   \and A.~V\"ogler\inst{4}
          }

   \institute{Astrophysics Group, Imperial College London, SW7 2AZ, U.K.
   \and Max-Planck-Institut f\"ur Sonnensystemforschung, Max-Planck-Strasse 2, D-37191 Katlenburg-Lindau, Germany
    \and School of Space Research, Kyung Hee University, Yongin, Gyeonggi, 446-701, Korea   
    \and Sterrekundig Instituut, Utrecht University, Postbus 80 000, 3508 TA Utrecht, The Netherlands
    }    
   \date{Received date; accepted date}

% \abstract{}{}{}{}{} 
 
  \abstract
  % context heading (optional)
  {Changes in the solar surface area covered by small-scale magnetic elements are thought to cause long-term changes in the solar spectral irradiance, which are important for determining the impact on Earth's climate.}   
  % aims heading (mandatory)
   {To study the effect of small-scale magnetic elements on total and spectral irradiance, we derive their contrasts from 3-D MHD simulations of the solar atmosphere. Such calculations are necessary since measurements of small-scale flux tube contrasts are confined to a few wavelengths and suffer from scattered light and instrument defocus, even for space observations.}
  % methods heading (mandatory)
 {To test the contrast calculations, we compare rms contrasts from simulations with those obtained with the  broad-band filter imager mounted on the Solar Optical Telescope (SOT) onboard the Hinode satellite and also analyse centre-to-limb  variations (CLV). The 3-D MHD simulations include the interaction between convection and magnetic flux tubes. They have been run with non-grey radiative transfer using the \textit{MURaM} code.  Simulations have an average vertical magnetic field of 0G, 50G, and 200G. Emergent intensities are calculated with the spectral synthesis code ATLAS9 and are convolved with a theoretical point-spread function to account for the properties of the  observations' optical system.}
  % results heading (mandatory)
 {We find reasonable agreement for simulated and observed intensity distributions in the visible continuum bands. Agreement is poorer for the CN and G-Bands. The analysis of the simulations exhibits a potentially more realistic centre-to-limb behaviour than calculations based on 1-D model atmospheres.}
  % conclusions heading (optional), leave it empty if necessary 
  {We conclude that starting from 3-D MHD simulations represents a powerful approach to obtaining intensity contrasts for a wide wavelength coverage and for different positions on the solar disk. This also paves the way for future calculations of facular and network contrast as a function of magnetic fluxes.}


   \maketitle

 \begin{figure}
\centering
\begin{picture}(200,170)
\put(-70,-180){\begin{picture}(0,0) \includegraphics{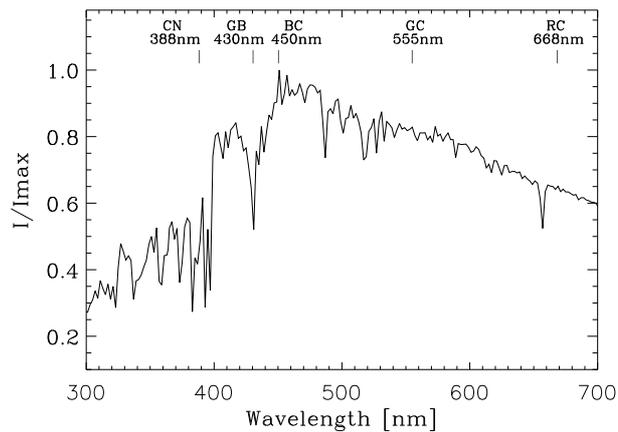} \end{picture}}
\end{picture}
\caption{Spectrum of the visible wavelength regions at disk centre (calculated with ATLAS9 for the non-magnetic simulation) with the positions of the five Hinode filters indicated.}
\label{fig:spectrum}
\end{figure}

%%%%%%%%%%%%%%%%%%%%%%%%%%%%%%%%%%%%%%%%%%%%%%%%%%%%%%%%%%%%%%%%%
\section{Introduction}
%%%%%%%%%%%%%%%%%%%%%%%%%%%%%%%%%%%%%%%%%%%%%%%%%%%%%%%%%%%%%%%%%

 \begin{figure}
\centering
\begin{picture}(200,170)
\put(-70,-180){\begin{picture}(0,0) \includegraphics{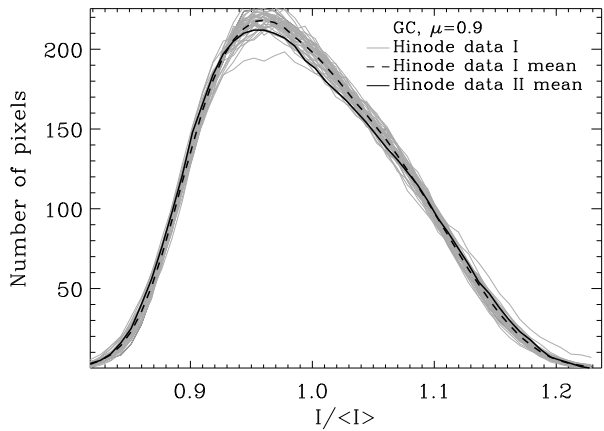} \end{picture}}
\end{picture}
\caption{Comparison of Hinode datasets I (Mercury transit) and II for the green continuum wavelength at the limb angle   $\mu=0.9$. For dataset I, we plot 37 histograms where each includes 10 random extracts of the whole observed image (grey lines) and the mean histogram (solid line). We overplot the mean of the analogous quantity for dataset II (black dashed line).}
\label{fig:comphinodes}
\end{figure}

\noindent Understanding  the variability of the Earth's atmosphere requires the ability to distinguish between anthropogenic and natural drivers of climate change. One important natural driver influencing our climate is solar irradiance (i.e. the  radiative flux of the Sun, measured above the Earth's atmosphere) and in particular its variation. Changes in the Sun's total irradiance influence the amount of energy reaching the Earth. In addition, changes in the solar spectral irradiance, i.e. the changes in the Sun's brightness at a certain wavelength,  influence the chemistry of the Earth's upper atmosphere \citep[e.g.][]{haigh2007}.  Finally, the heliospheric magnetic field  \citep{potgieter1998, simpson1998}  shields the Earth from galactic cosmic rays which have been proposed to increase cloud cover \citep{marschsvensmark2000}. The solar surface magnetic field, which manifests itself in, e.g., sunspots, faculae, and the network is able to reproduce solar irradiance variations on solar-cycle time scales \citep[e.g.][]{krivovaetal2003, wenzleretal2005, wenzleretal2006}. For longer time scales, however, there is considerable uncertainty. While sunspots and faculae dominate the solar irradiance changes on timescales of days to weeks, the small-scale magnetic elements in the network are considered to be the main cause of long-term changes in the solar irradiance \citep{foukallean1988, solankifligge2002}. The contribution from sunspots to irradiance changes is well understood and produces a decrease in solar irradiance. The contribution from faculae and mainly from smaller-scale magnetic elements composing the photospheric network increases the solar irradiance and is still being investigated. In particular, the spectral response is poorly understood. For a better understanding of irradiance variations, it is important to study the centre-to-limb variation and the wavelength dependence of the emergent intensity.

The direct measurement of the solar network contrast to determine the impact of these small-scale magnetic elements remains difficult, since it is (even for space instruments) affected by the instrument's aperture, scattered light, instrument defocus, etc. and additionally limited to a few wavelengths. Therefore, we calculate emergent intensities from 3-D simulations of solar magneto-convection and calculate the intensity variation between brighter and darker features to compare the simulations with observations.  Filtergrams with a high and constant spatial resolution  can be obtained with the broad-band filter imager (BFI) mounted on the Solar Optical Telescope (SOT, \cite{tsunetaetal2008, ichimotoetal2008, suematsuetal2008, shimizuetal2008}) onboard the HINODE spacecraft \citep{kosugietal2007}. For a comparison of measured and simulated granulation contrasts, we convolved the simulations with the point-spread function (PSF)  of the observation's optical system \citep{mathewetal2009} to take into account its effects. Here, we analyse simulations with different average magnetic field, while previous studies mainly included nonmagnetic simulations:  \cite{danilovicetal2008} analysed Hinode/SP data and compared the intensity contrast of a Hinode/SP continuum map of the quiet Sun with predictions of nonmagnetic simulations, and \cite{wedemeyervoort2009} discussed Hinode/SOT observations in comparison to nonmagnetic simulations to determine the continuum intensity distributions of the solar photosphere.

The paper is organised as follows: the observations are described in Sect.~2, the simulations, the intensity synthesis originating from them and the subsequent convolution are introduced in Sect.~3.  We compare the Hinode/SOT/BFI  observations with the simulations in Sect.~4 (in terms of histograms and rms contrasts) and also discuss the statistical aspects of this comparison. The centre-to-limb behaviour of the intensity contrast in our simulations is analysed in Sect.~5.  

\begin{figure*}
\centering
\begin{picture}(250,350)
\put(-140,145){\begin{picture}(0,0) \includegraphics{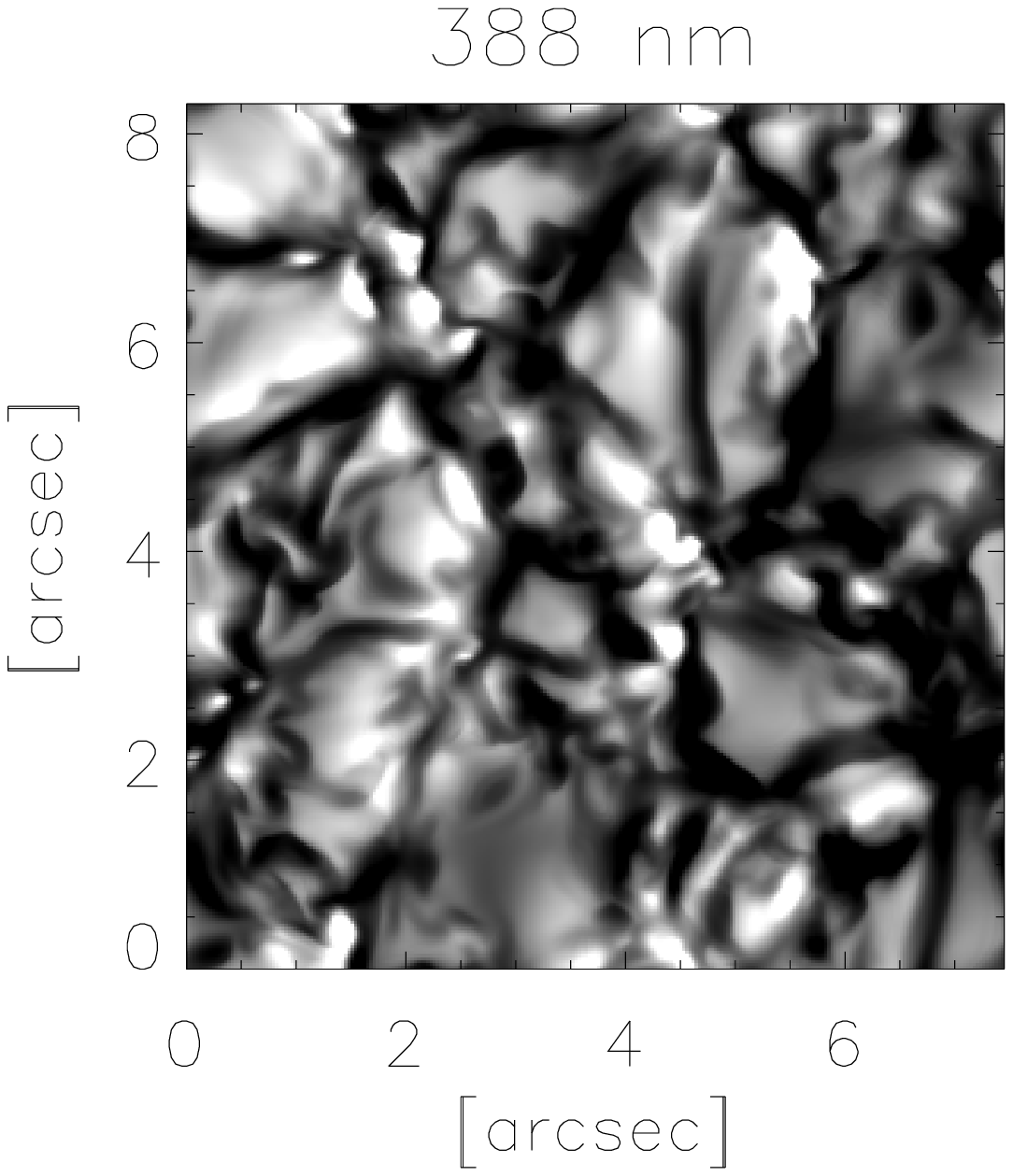} \end{picture}}
\put(-40,145){\begin{picture}(0,0) \includegraphics{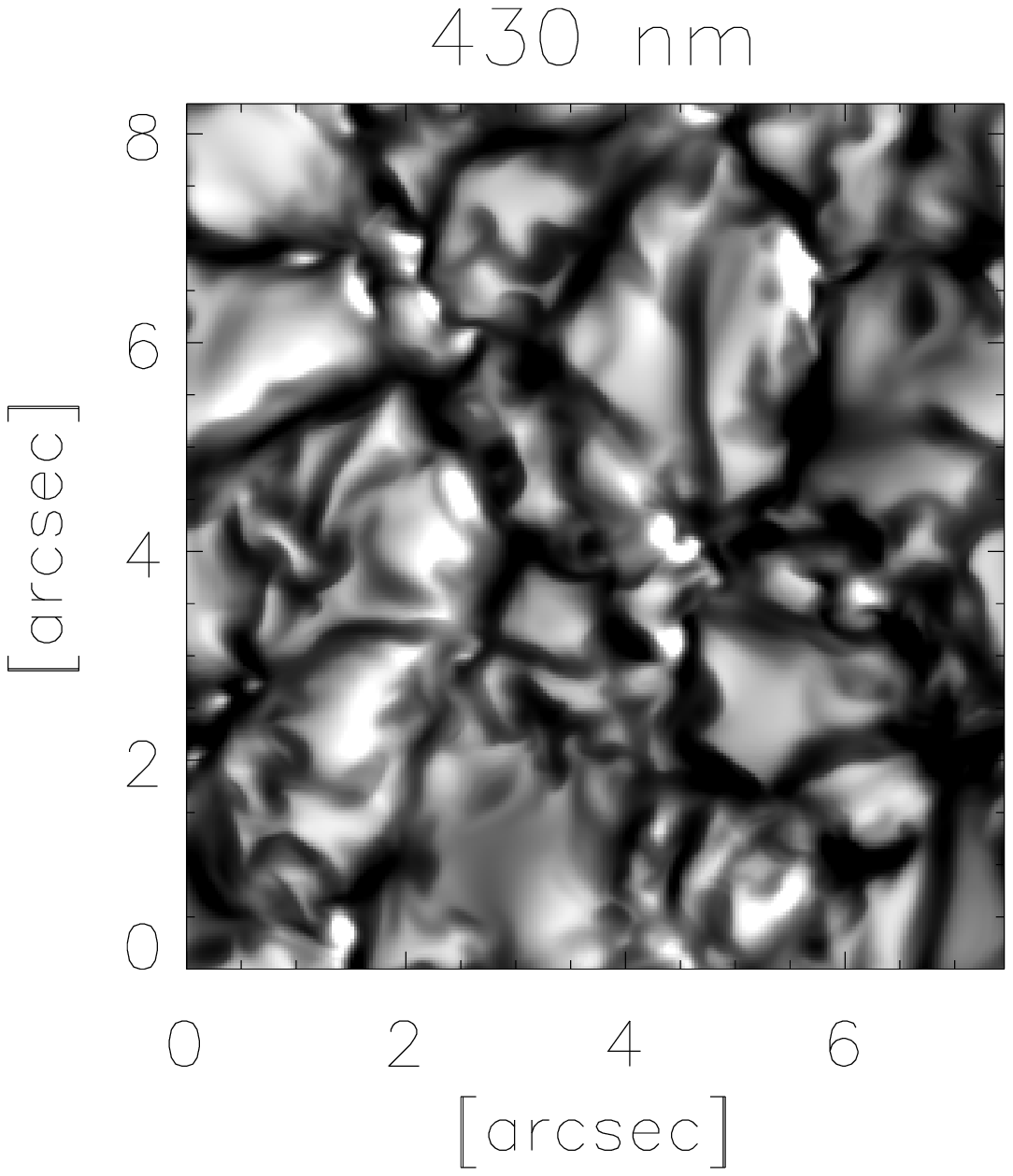} \end{picture}}
\put(60,145){\begin{picture}(0,0) \includegraphics{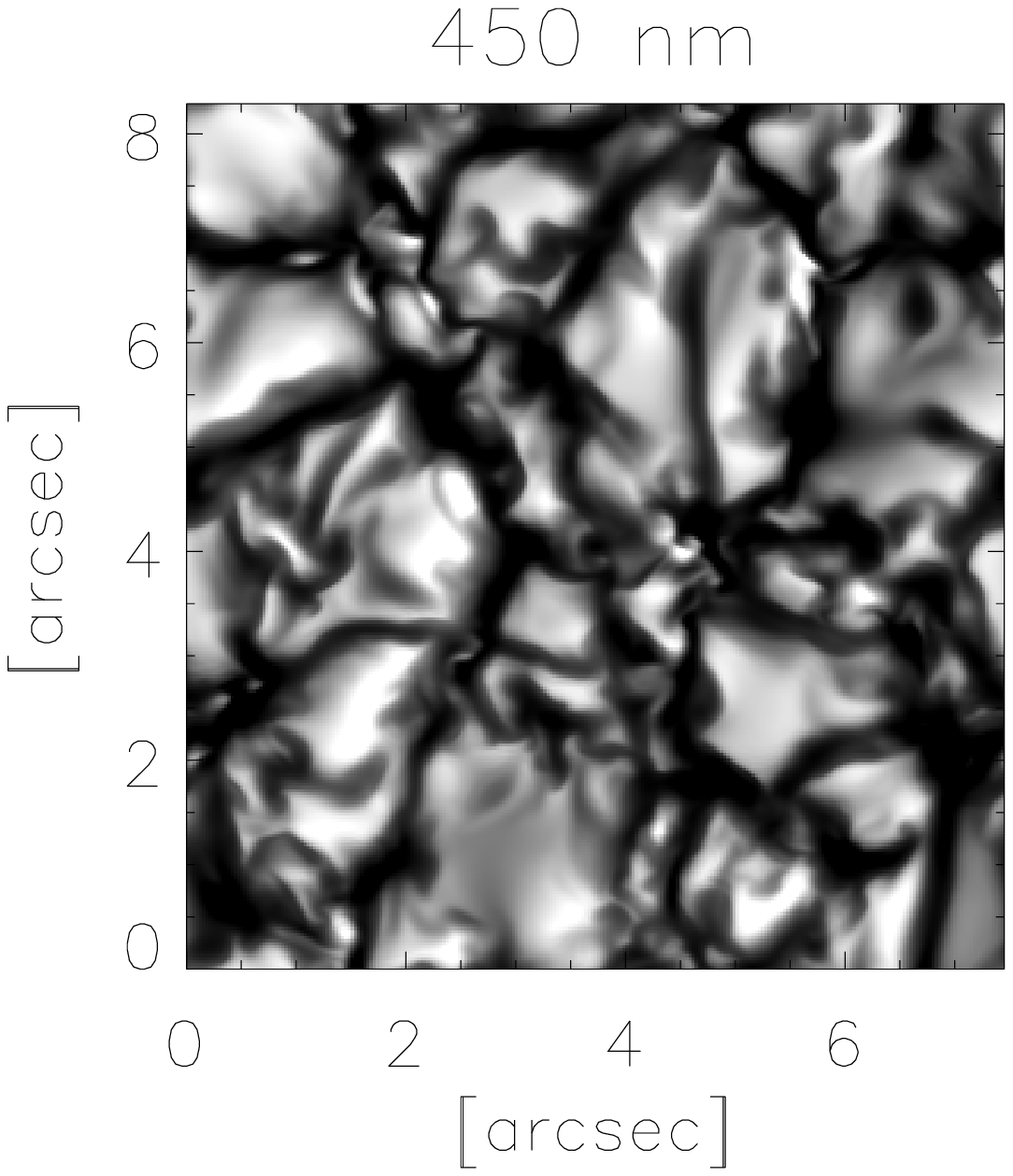} \end{picture}}  
 \put(160,145){\begin{picture}(0,0) \includegraphics{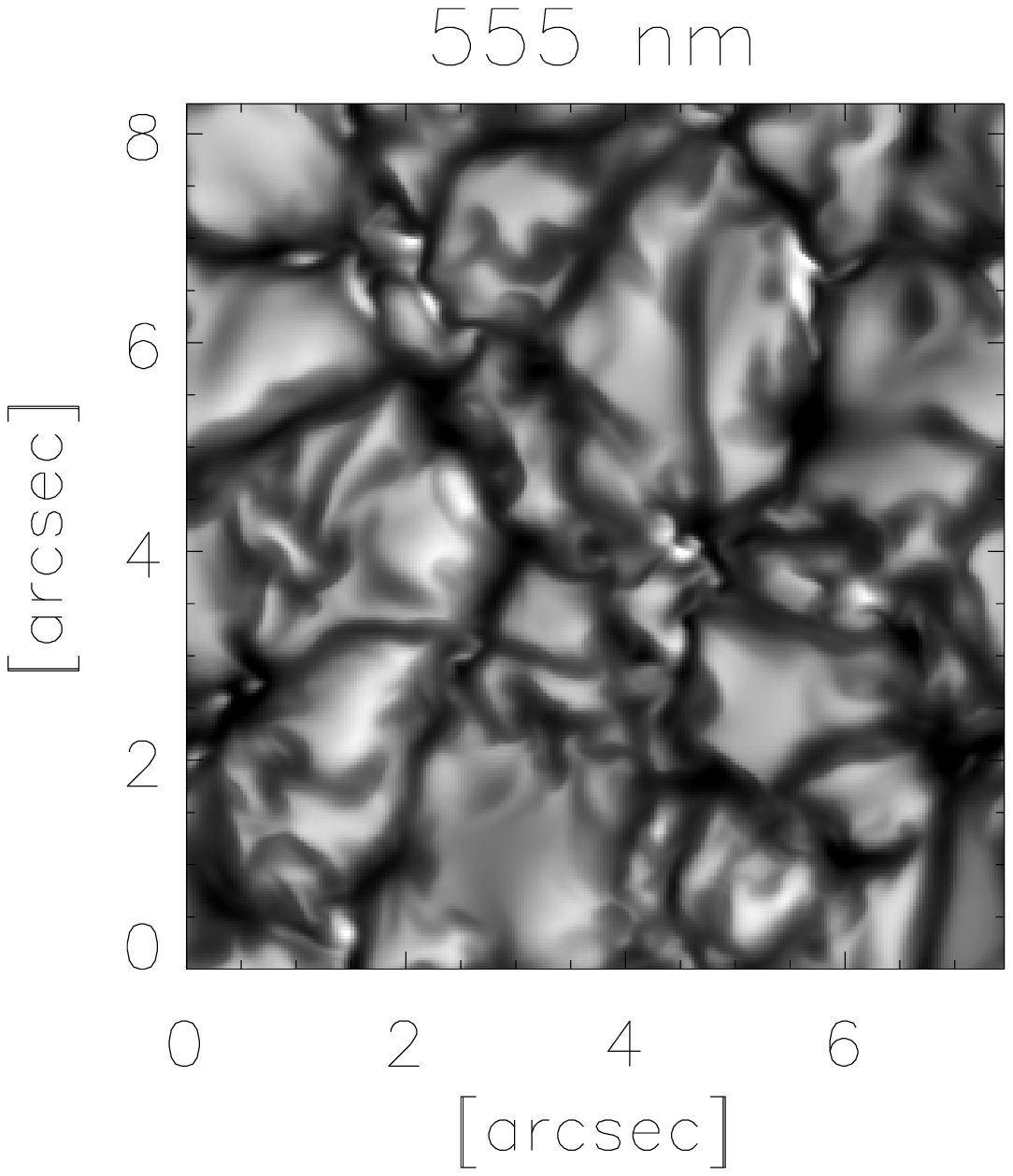} \end{picture}}
\put(260,145){\begin{picture}(0,0) \includegraphics{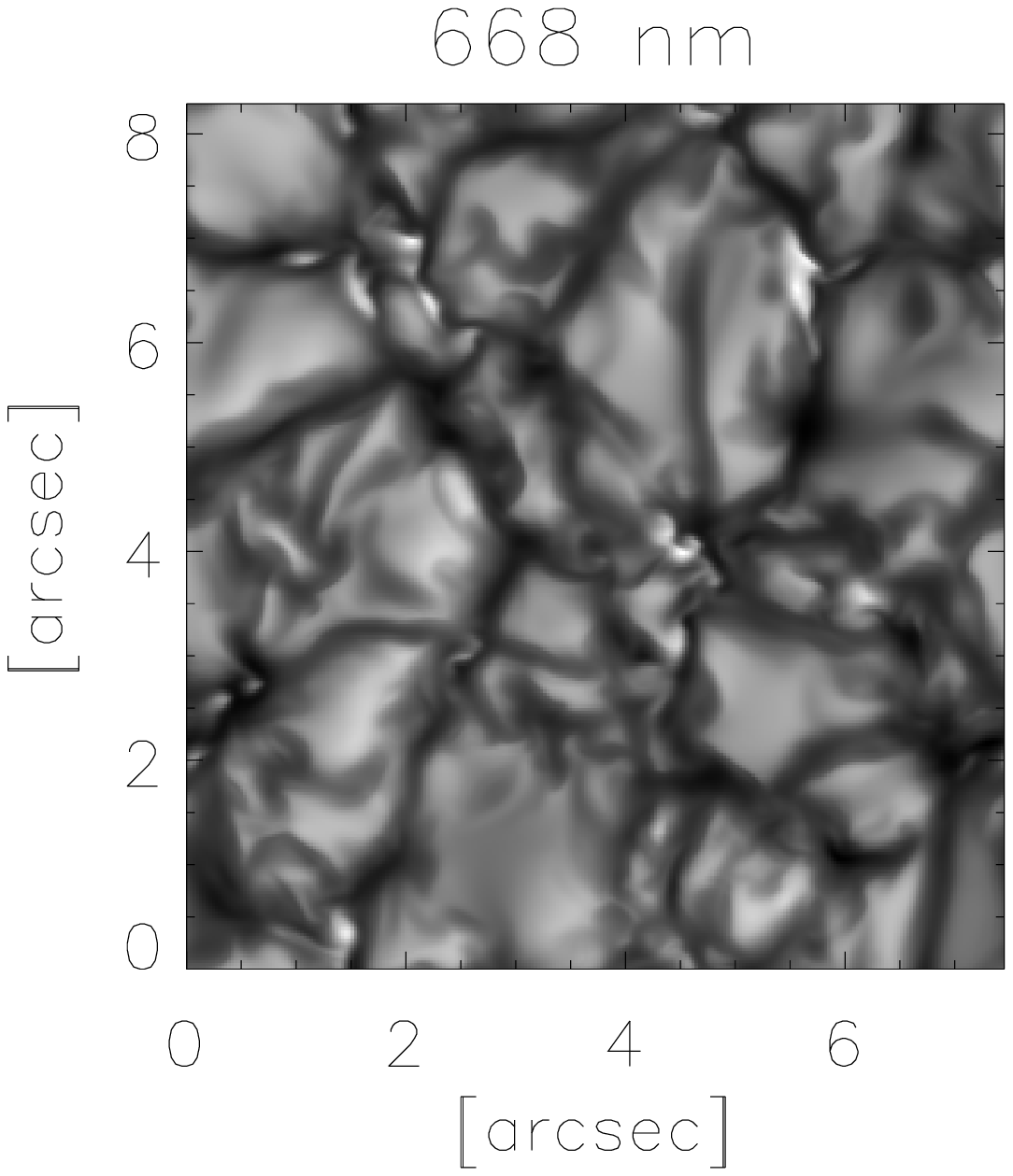} \end{picture}}  
\put(-140,30){\begin{picture}(0,0) \includegraphics{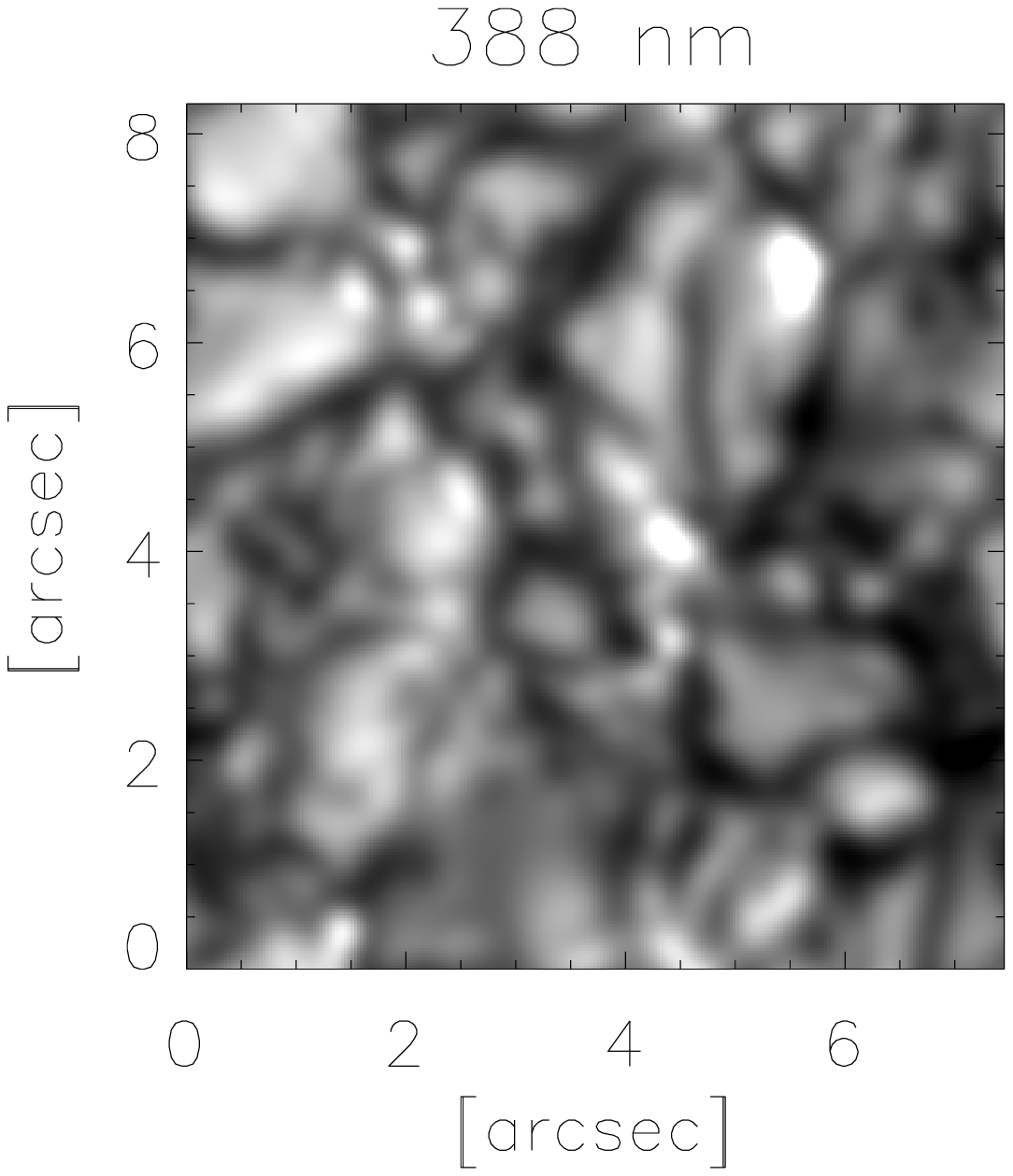} \end{picture}}
\put(-40,30){\begin{picture}(0,0) \includegraphics{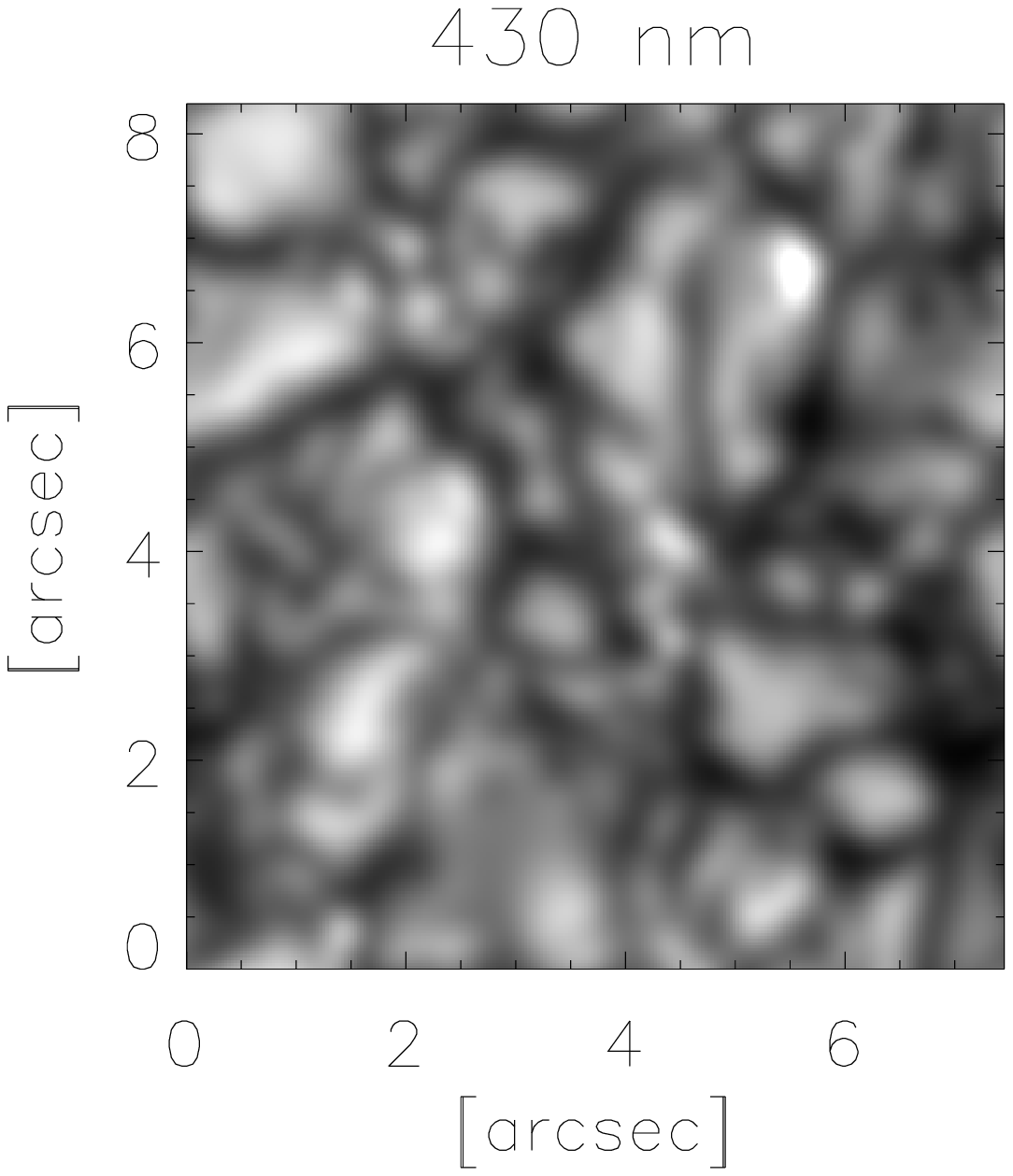} \end{picture}}
\put(60,30){\begin{picture}(0,0) \includegraphics{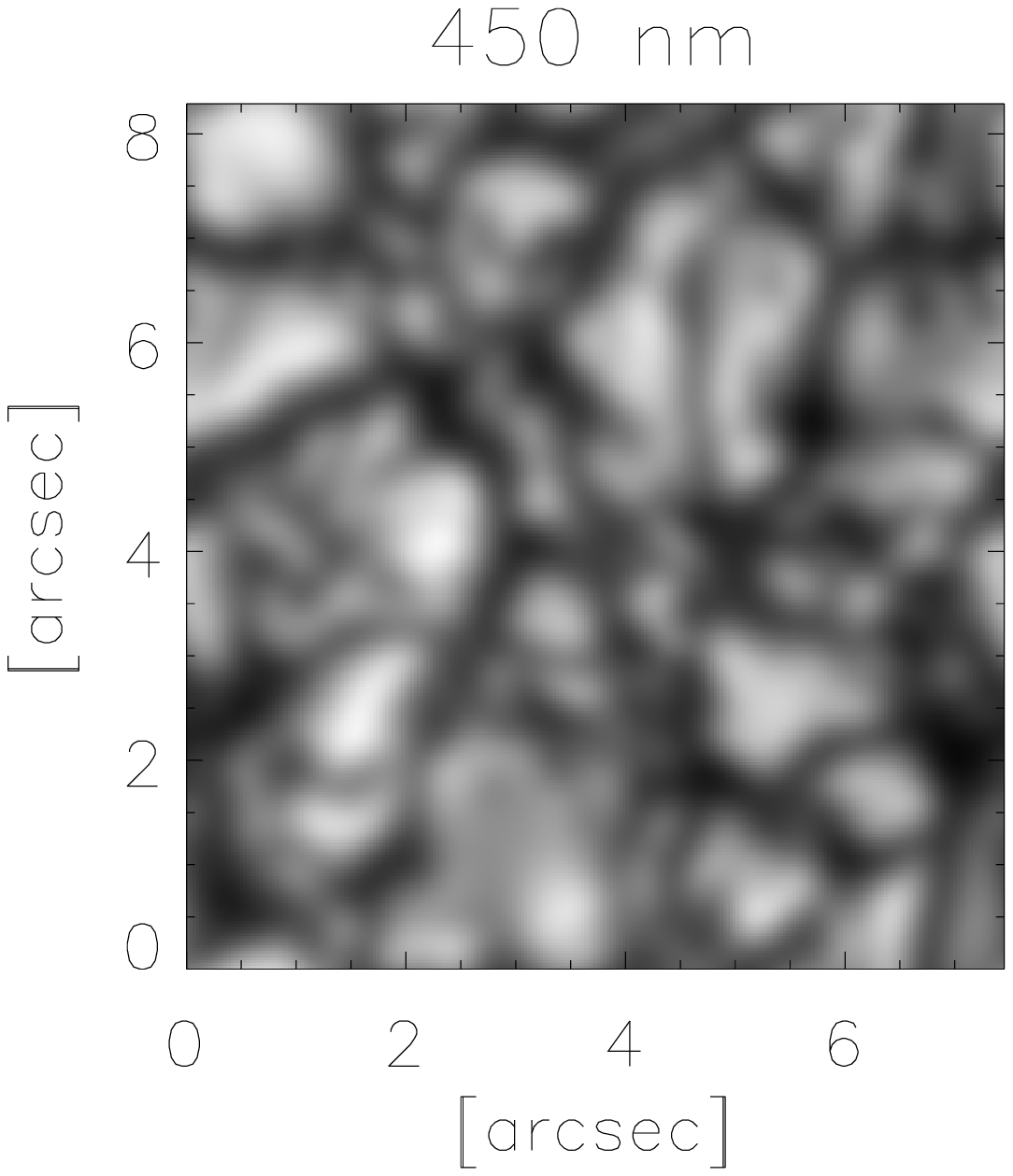} \end{picture}}  
 \put(160,30){\begin{picture}(0,0) \includegraphics{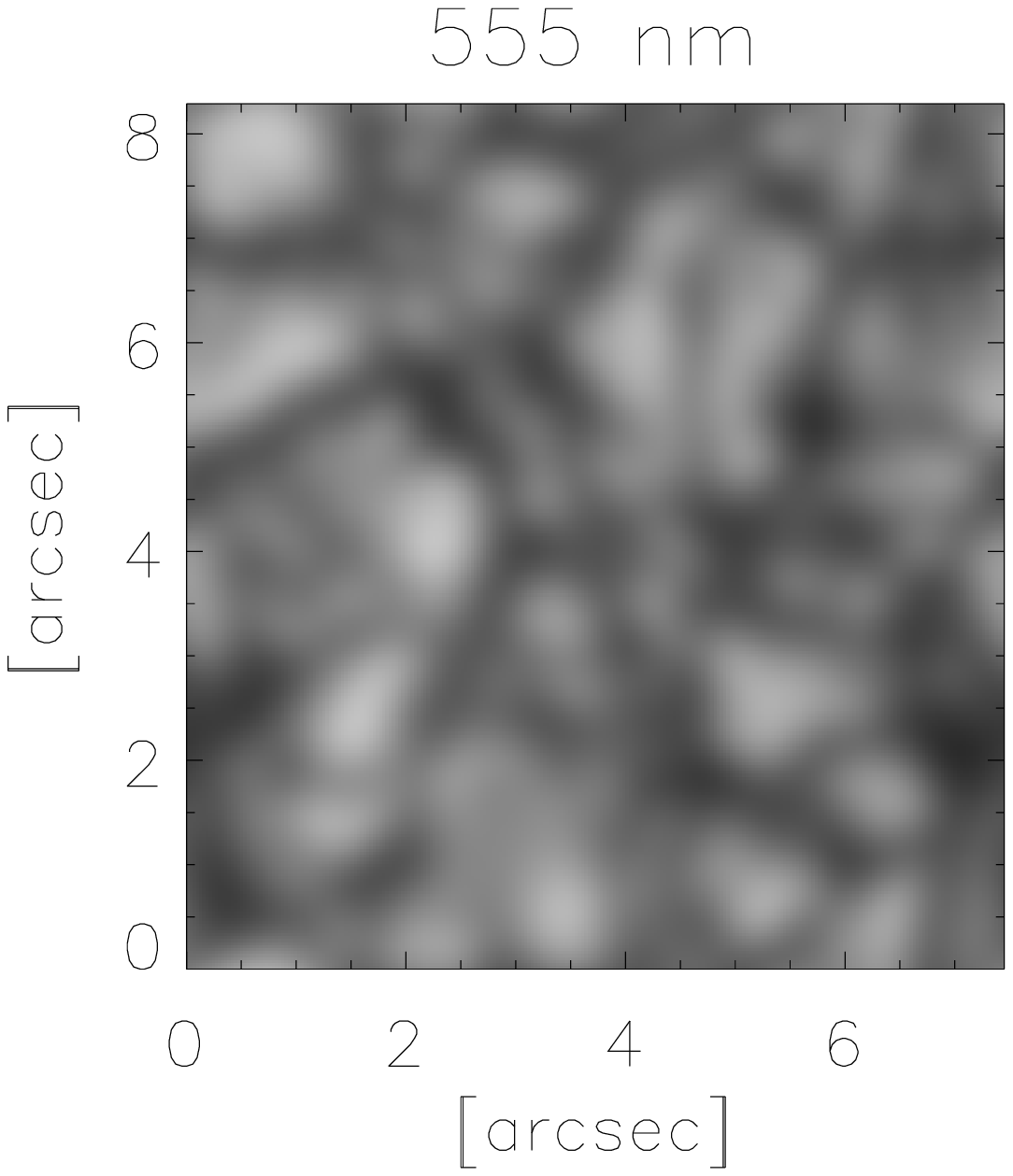} \end{picture}}
\put(260,30){\begin{picture}(0,0) \includegraphics{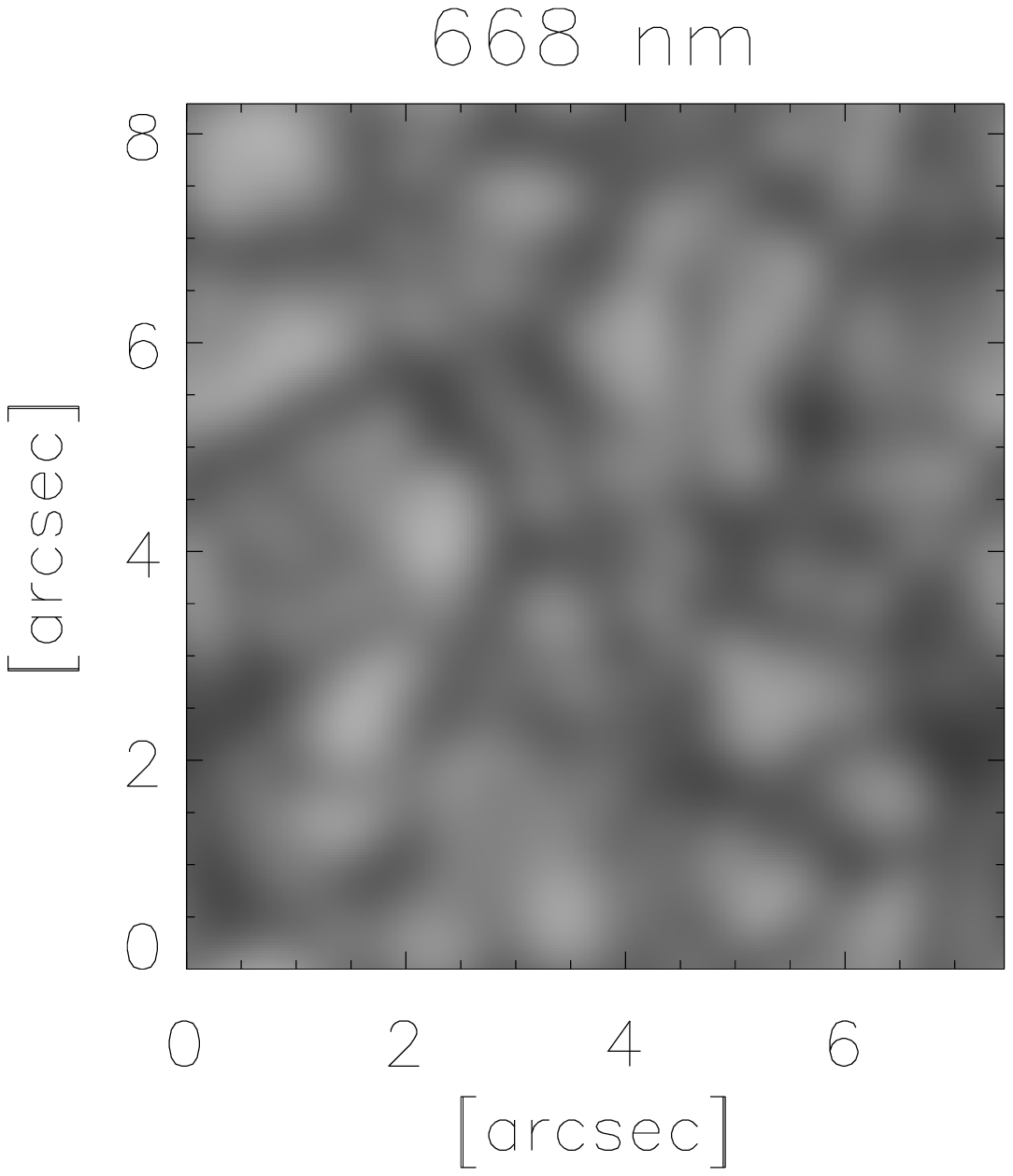} \end{picture}}  
\put(-140,-85){\begin{picture}(0,0) \includegraphics{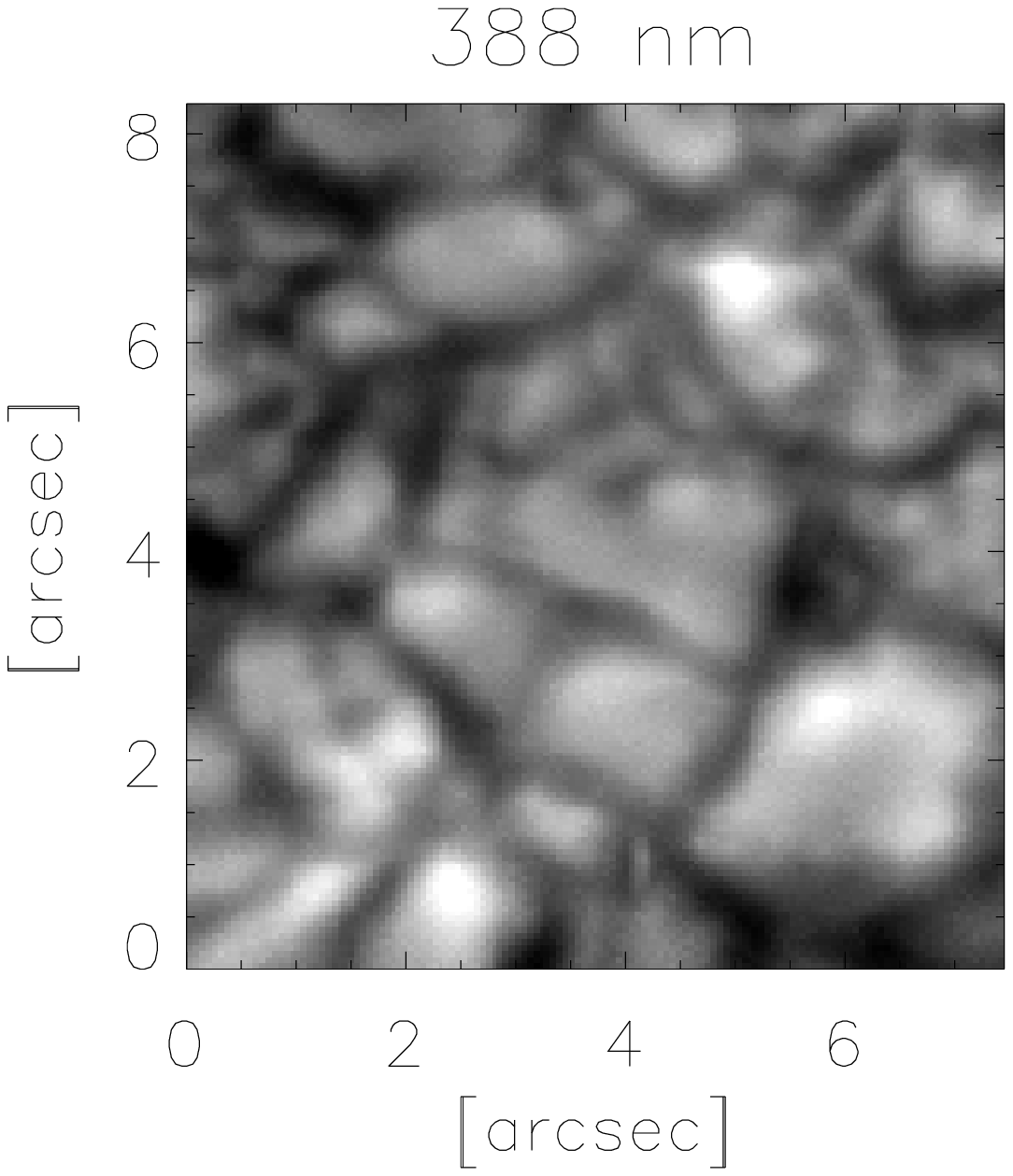} \end{picture}}
\put(-40,-85){\begin{picture}(0,0) \includegraphics{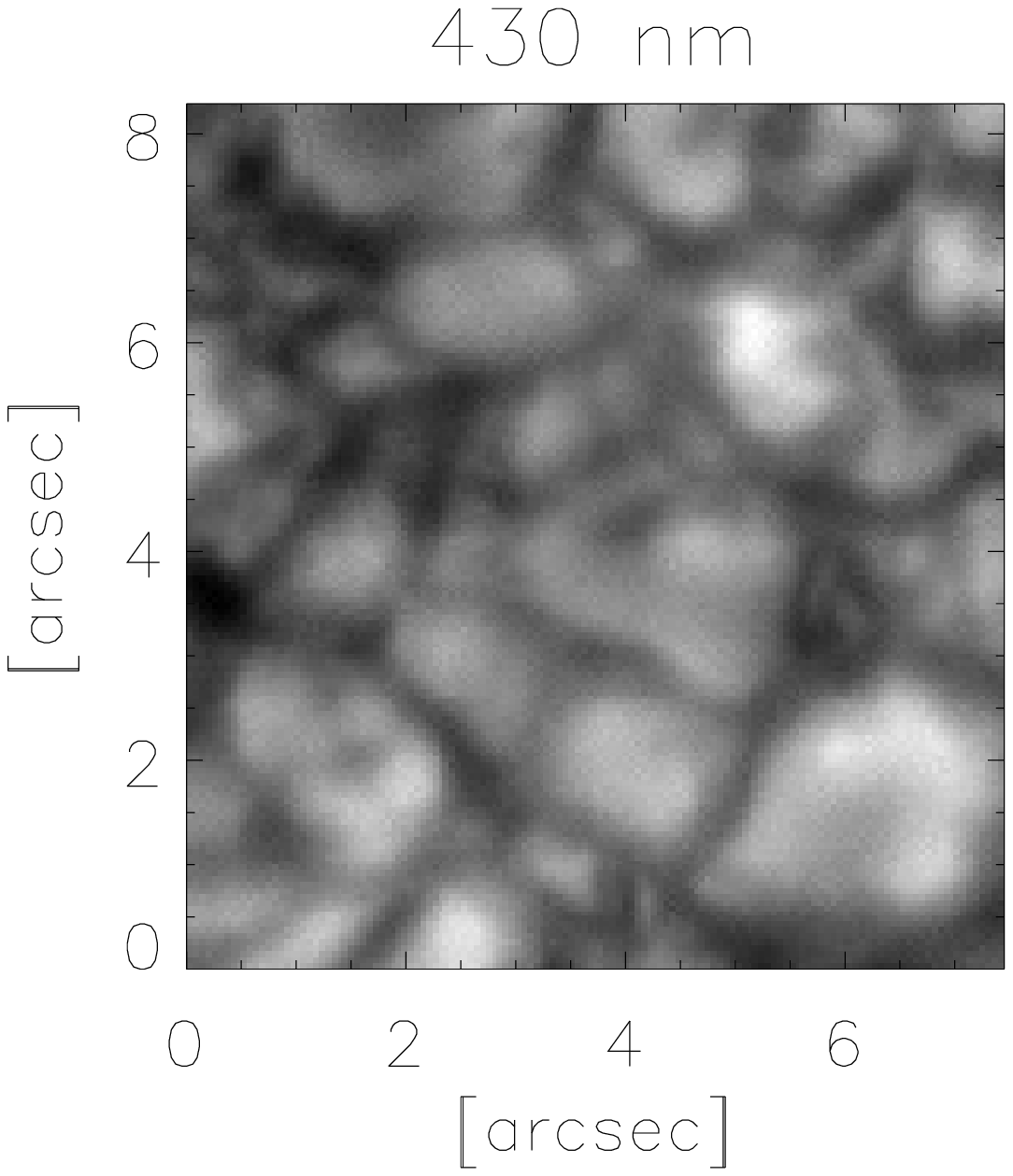} \end{picture}}
\put(60,-85){\begin{picture}(0,0) \includegraphics{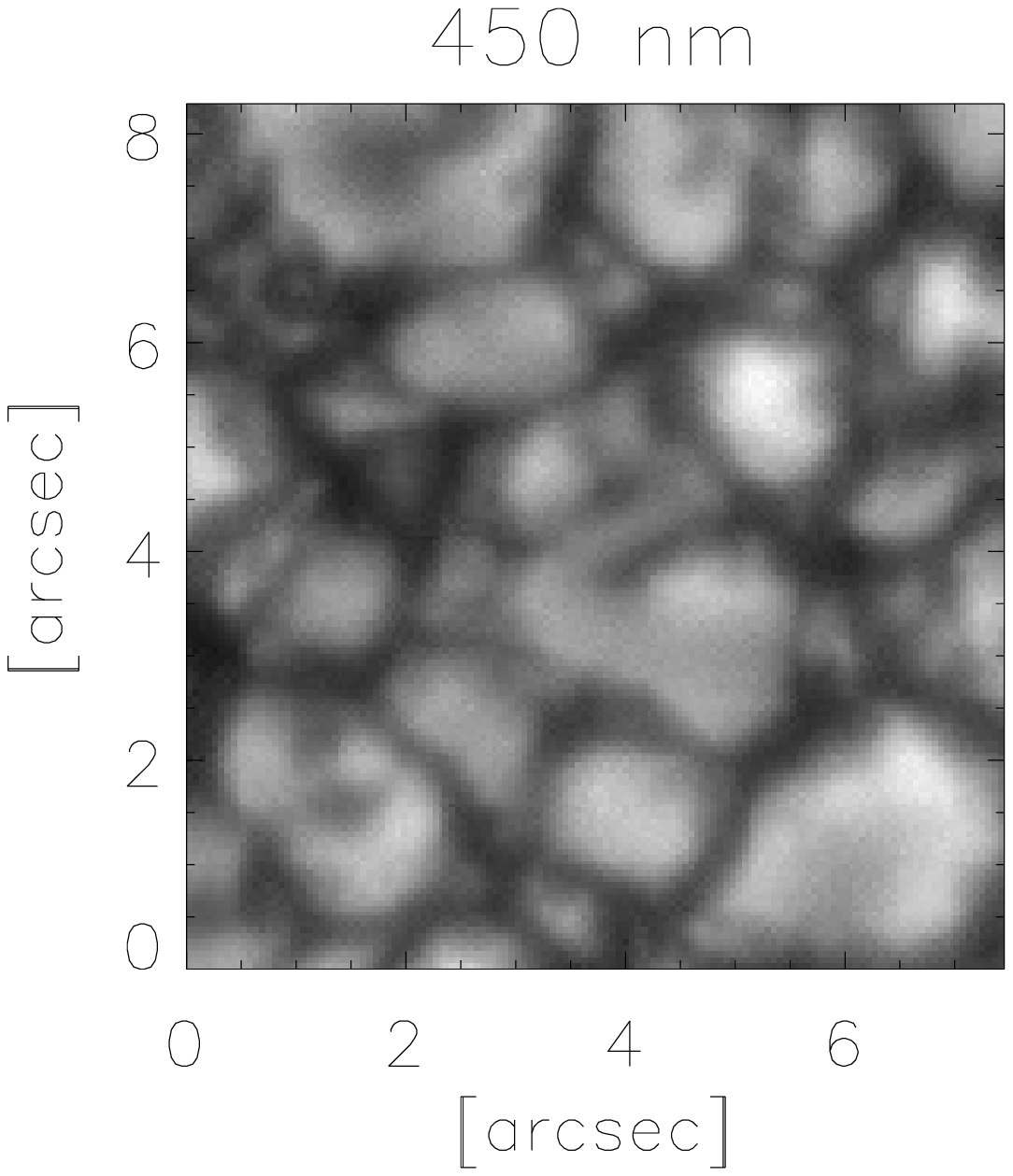} \end{picture}}  
 \put(160,-85){\begin{picture}(0,0) \includegraphics{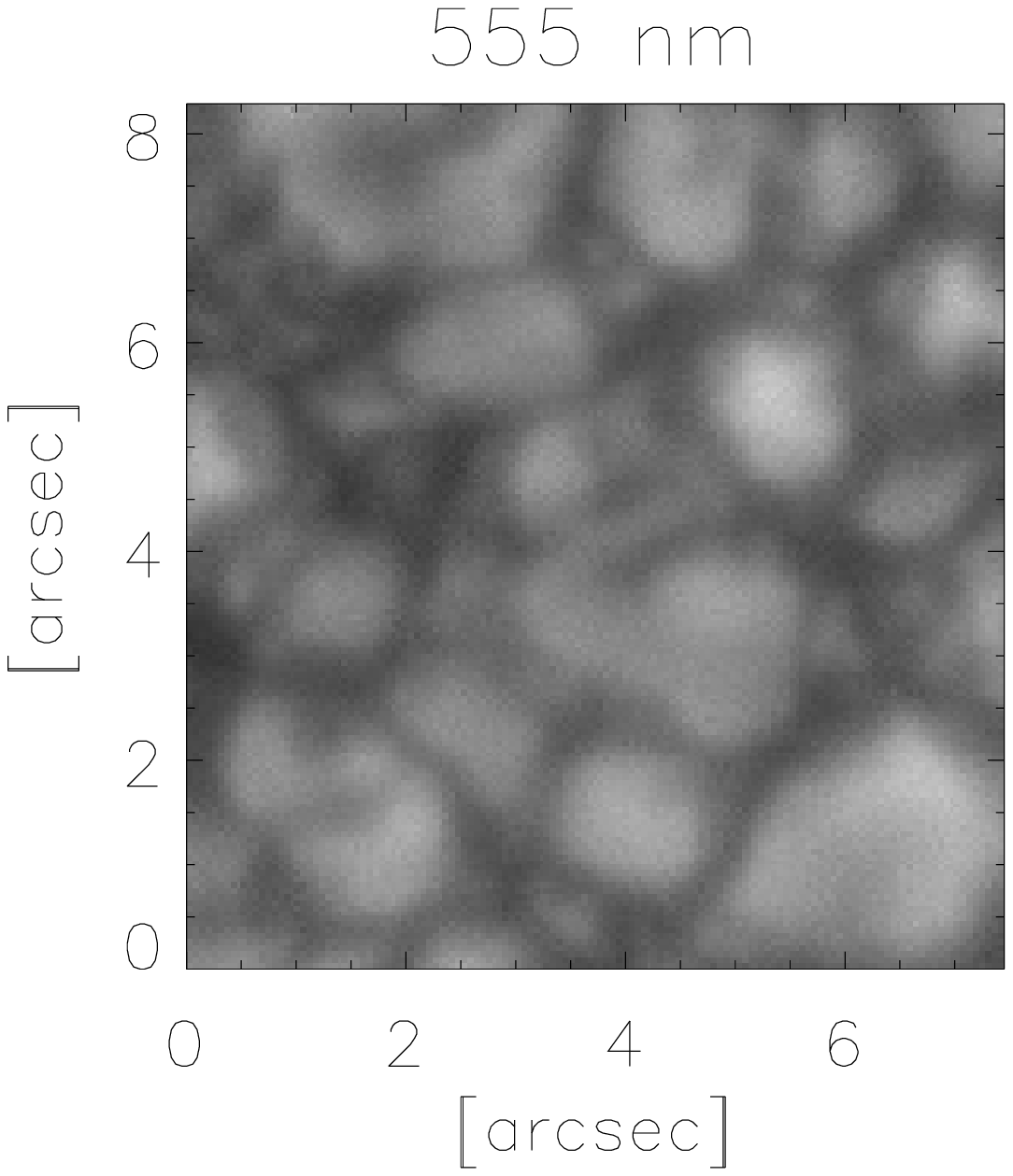} \end{picture}}
\put(260,-85){\begin{picture}(0,0) \includegraphics{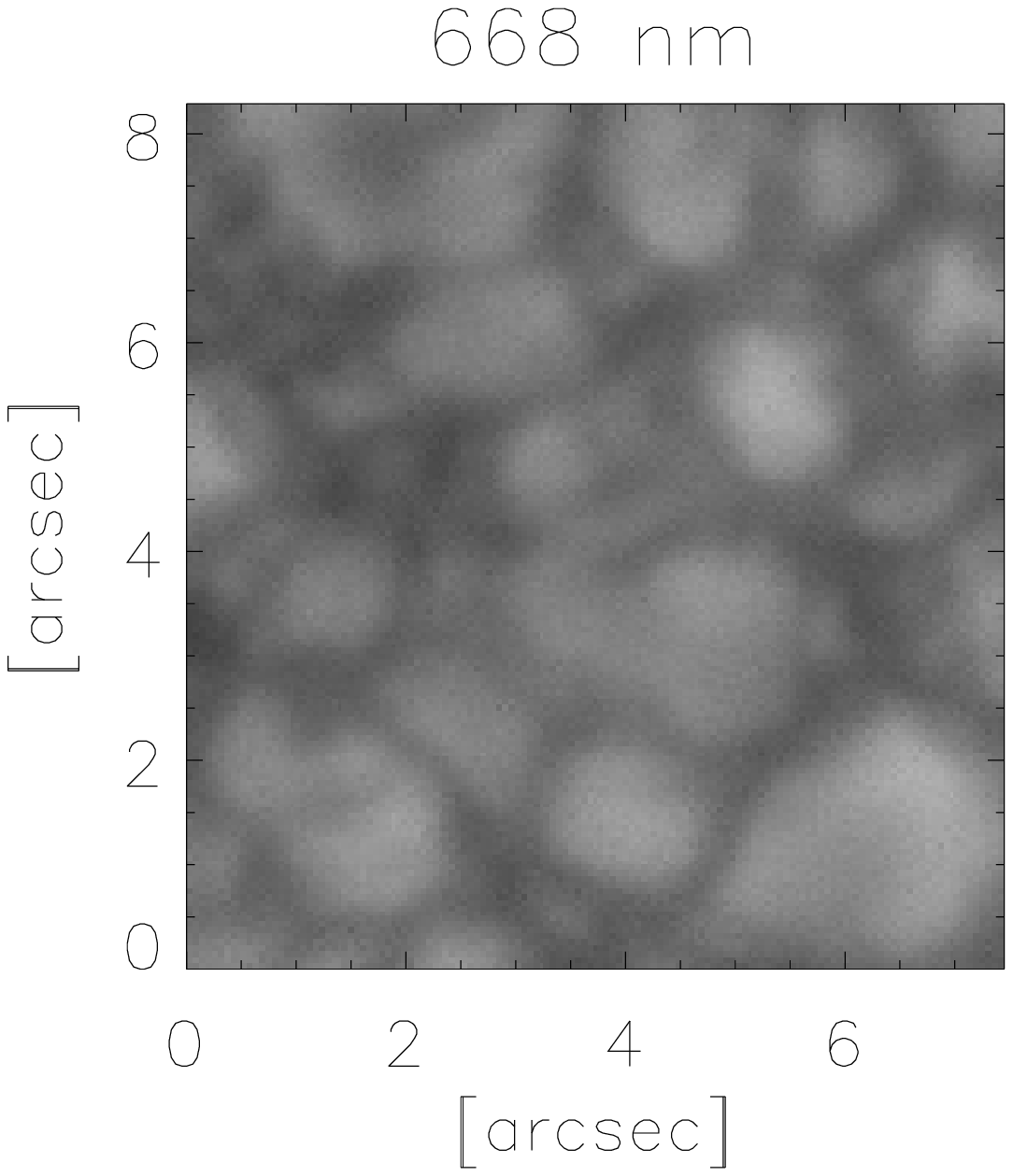} \end{picture}}  
\end{picture}
\caption{Simulations (original and convolved) and observations in the 5 Hinode/SOT filters. \textit{Top:} Original simulation snapshots with an input magnetic field of 50G. Black: 52\% below the mean, white: 68\% above the mean value. \textit{Middle:} Same simulation snapshots convolved with an appropriate point-spread function (PSF). \textit{Bottom:} Extracts from observed image. The black/white scale for the convolved simulations and the observations is such that  black is 33\% below the mean, and  white is 40\% above the mean value.}
\label{fig:muram50G_hinode_unconv_conv_all}
\end{figure*}

\section{Observations}
%%%%%%%%%%%%%%%%%%%%%%%%%%%%%%%%%%%%%%%%%%%%%%%%%%%%%%%%%%%%%%%%%

In this work we consider 2 different datasets from the Hinode Solar Optical Telescope (SOT):

\textbf{Dataset I} was recorded during the Mercury transit on 8th Nov  2006 (21:38:31-21:39:16 UT) in 5 wavelength filters: CN bandhead ('CN', 388.3 nm), G band ('GB', 430.5 nm), blue continuum ('BC', 450.4 nm), green continuum ('GC', 555.0 nm) and red continuum ('RC', 668.4 nm) (Fig.~\ref{fig:spectrum}). The observed region corresponded to quiet Sun. The full observed Hinode image covers approximately 223 $\times$ 112 arcsec and the pixel size is 0.05 arcsec. The heliocentric coordinates at the centre of the image were (0,-419), corresponding to a limb angle of $\mu=0.9$, where cos $\theta=\mu$, and $\theta$ is the the angle between the line of sight and the line through the centre of the Sun.  The dark correction and flat fielding were done with SSW-IDL routines. The Mercury transit  allowed the calculation of the point-spread function (PSF) with high confidence.    Details of the calculation of the PSF and of this particular data set can be found in \cite{mathewetal2009}.

 \begin{figure}
\centering
\begin{picture}(200,170)
\put(-70,-180){\begin{picture}(0,0) \includegraphics{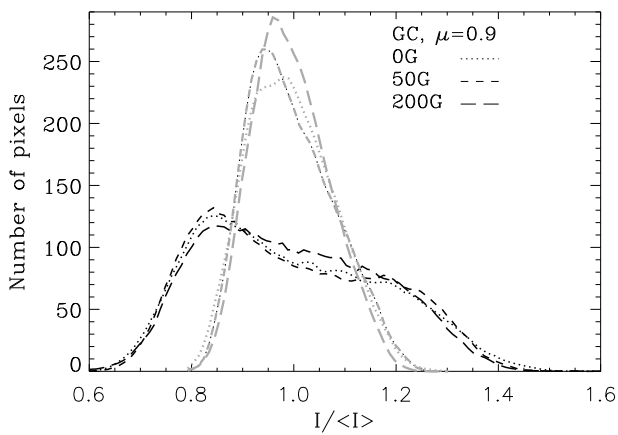} \end{picture}}
\end{picture}
\caption{Histograms of the intensity distribution in the original simulation for the GC wavelength at the limb angle   $\mu=0.9$: 0G (black dotted), 50G (black dashed), 200G (black long dashes). We overplot (in grey) the corresponding histograms for the convolved simulation  to illustrate the effect of convolving with the PSF.}
\label{fig:hist_gc_unconv}
\end{figure}
%as shown in Figure \ref{fig:muram_hinode_conv_histo_all}

\textbf{Dataset II} was recorded on 8th March 2007 (18:02:32-20:51:46 UT)  in the same wavelength filters as dataset I. While quiet Sun was observed also in this observation, the binning was different with a  pixel size of 0.10 arcsec. The data were recorded in 12 steps from the east to the west limb, resulting in the availability of data at all limb angles. At each step, the observed image covers approximately a field of view of  223 $\times$ 112 arcsec, as in dataset I.  For each image, we chose extracts such that their area corresponds to the desired limb angle solely. This allowed us to use dataset II for comparisons with calculations at limb angles from disk centre to the limb. It should be noted that we are observing different areas on the Sun at different limb angles and that their activity levels might be different.

Dataset I  was compared with dataset II  at $\mu=0.9$ to check the validity of the PSF for dataset II. An example is shown in  Fig.~\ref{fig:comphinodes}, where we compare histograms for the two datasets in the green continuum (GC). The number of resulting subimages in the two datasets are naturally different: For dataset I, we segmented the observed image  into 373 sub-images (extracts) of the desired size.  Four extracts had to be excluded due to the Mercury transit. To illustrate the spread in dataset I, we overplot in Fig.~\ref{fig:comphinodes} (in grey) 37 histograms where each includes 10 random extracts of the whole observed image. For dataset II, we have 151 subimages, and, for ease of comparison, we confine ourselves to plotting only the mean histogram over all the 151 histograms of the subimages. For dataset I, the subimages are scaled to the pixelsize of subimages in dataset II, which, for the limb angle $\mu=0.9$ implies 68 $\times$ 76 pixels per subimage. The mean of dataset II appears to be a bit wider than that of dataset I, but overall we find a good agreement. We follow the procedure outlined in Numerical Recipes \citep{pressetal} and calculate the $\chi^2$ statistic for two binned distributions (with unequal numbers of data points), where $R_{i}$ is the number of events in bin $i$ for the first data set, and $S_{i}$ the number of events in the same bin $i$ for the second data set:

\begin{equation}\label{eq:chisquare}
\hspace{1cm} \chi^2=\sum_{i} \frac{\left(R_{i}\sqrt{\frac{S}{R}}-{S_{i}\sqrt{\frac{R}{S}}}\right)^2}{R_{i}+S_{i}},
\end{equation} where 

\begin{equation}\label{eq:ri}
\hspace{1cm} R=\sum_{i} R_{i} \hspace{0.5cm} and \hspace{0.5cm}  S=\sum_{i} S_{i}
\end{equation} are the respective numbers of data points.

The  reduced $\chi^2$ value (i.e. the $\chi^2$ value divided by the number of degrees of freedom) between the mean of dataset I and dataset II is 0.42, suggesting that they are equivalent within the scatter. With similar results obtained for the other wavelength bands, this justifies the use of the same PSF for both datasets. 

For the comparison with simulations (described in the next section), an extract (or subimage) of an observation in this work is chosen such that it has the same physical size as one simulation snapshot, which corresponds to approximately 7.5 arcsec$\times$ 8.3 arcsec under a limb angle of $\mu=0.9$. Such an individual segment can then be directly compared with one simulation in terms of visual appearance, histograms, and contrast, as is done in Sect.~4.
% (see Fig.~\ref{fig:muram50G_hinode_unconv_conv_all}). To suppress image noise and in order to get a representative view against which to compare the simulations, we then averaged calculations (see Sect.~4.1) over all 373 extracts.

%%%%%%%%%%%%%%%%%%%%%%%%%%%%%%%%%%%%%%%%%%%%%%%%%%%%%%%%%%%%%%%%%
\section{Calculations}
%%%%%%%%%%%%%%%%%%%%%%%%%%%%%%%%%%%%%%%%%%%%%%%%%%%%%%%%%%%%%%%%%
\noindent \textbf{MHD simulations:} 
The 3-D MHD simulations have been run using the \textit{MURaM} code
 \citep{voegleretal2005} with non-grey radiative transfer using 4 opacity bins and 24 ray
 directions for determining the radiative heating/cooling rates \citep{voegleretal2004}. On the Sun, the computational box corresponds
horizontally to a square of 6 Mm $\times$ 6 Mm and has a vertical extension
of 1.4 Mm, covering the range between about 500 km above and 900 km
below the visible solar surface that is located at an average height of optical depth
unity. The cell size of the numerical grid is 20.8 km in the
horizontal directions and 14 km in the vertical. We used 3
simulation runs with initially homogeneous and unipolar
 vertical magnetic fields of 0G, 50G, and 200G, respectively. Since we wish to study the effect of
the surface magnetic field on the radiative output at the surface, we
do not prescribe the total radiative surface flux, but rather assume a
fixed constant entropy density of the fluid entering the simulation
box through the open lower boundary  \citep{voegleretal2005ii}. The value of the entropy density
has been chosen such that the resulting mean radiative energy flux in
the non-magnetic run matches the observed solar value to within
1\%. In all three simulations, the total radiative output at the top of the
 simulation box fluctuates by about $\pm 1$\% around its mean value in
 the course of the 5-minute oscillations in the box. Each simulation was run for about 150~min solar time after a
statistically stationary state of the magneto-convective system had
evolved. For the detailed analysis we then considered 10 equidistant
snapshots covering about 20~min in order to suppress
the effects of the 5-minute oscillations and to reduce the statistical
fluctuations. \cite{voegleretal2005ii} used these simulations to study the centre-to-limb variation of the bolometric intensity in dependence of the average vertical magnetic field.

\begin{figure*}
\centering
\begin{picture}(250,500)
\put(-130,290){\begin{picture}(0,0) \includegraphics{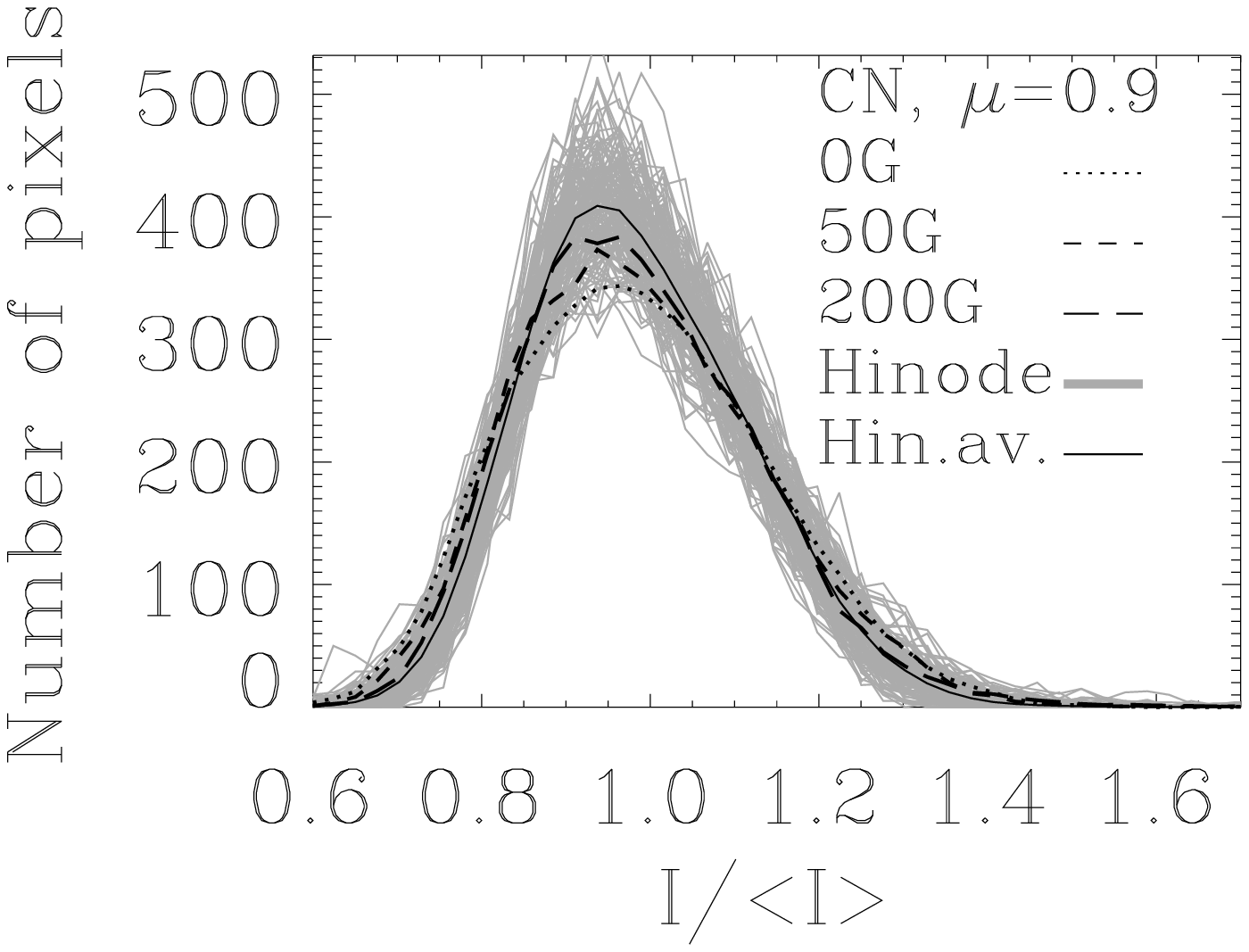} \end{picture}}
\put(-130,190){\begin{picture}(0,0) \includegraphics{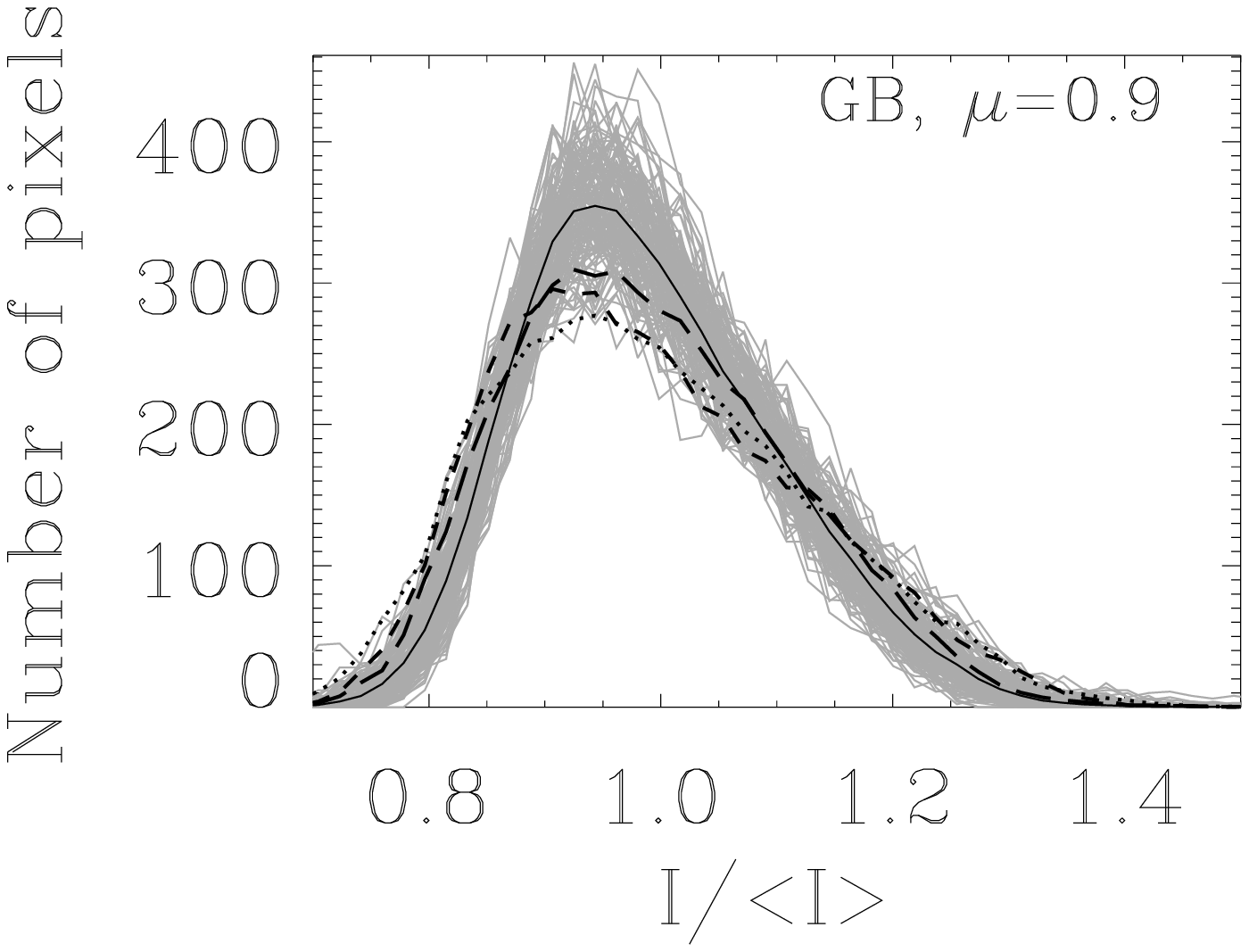} \end{picture}}
\put(-130,90){\begin{picture}(0,0) \includegraphics{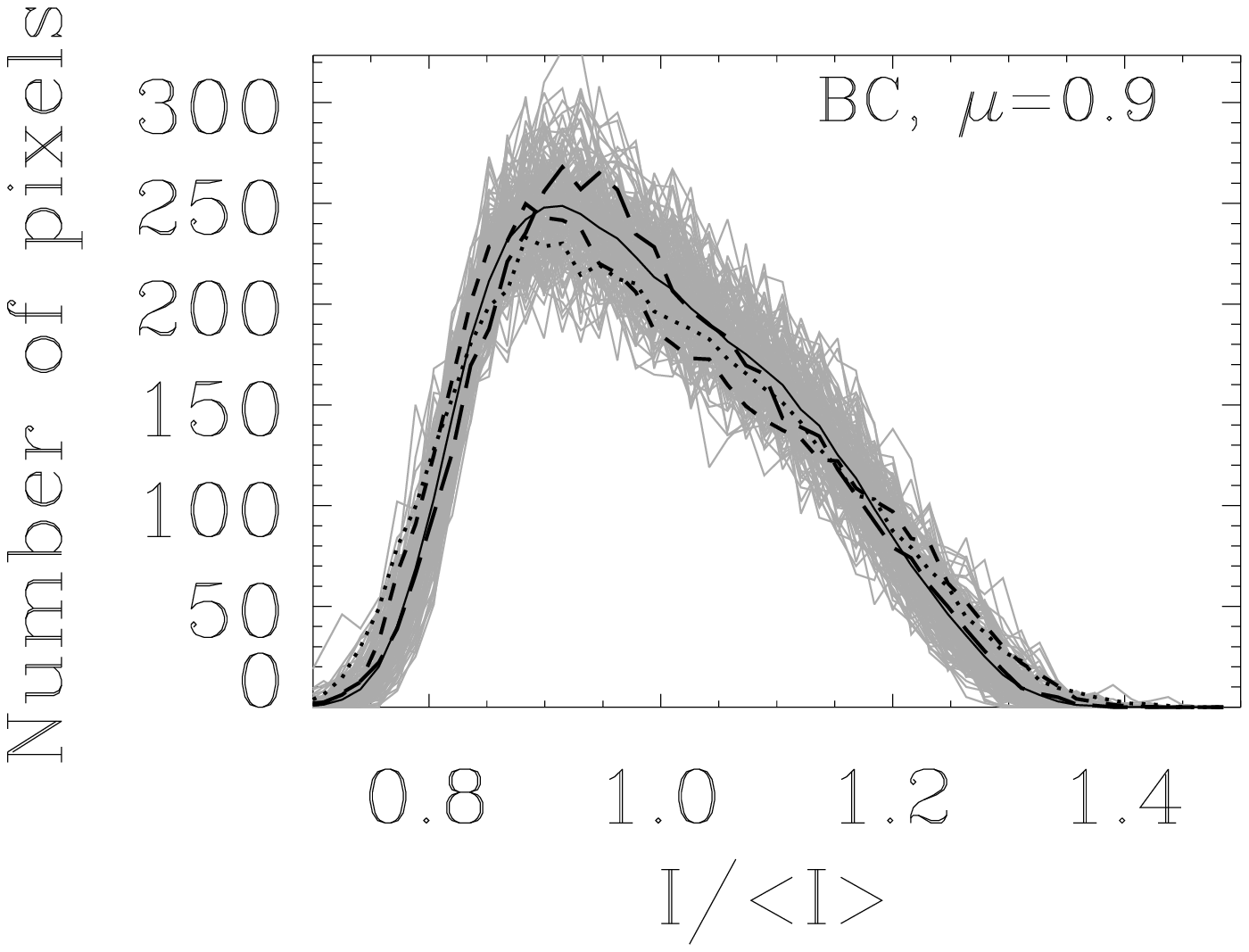} \end{picture}}
 \put(-130,-10){\begin{picture}(0,0) \includegraphics{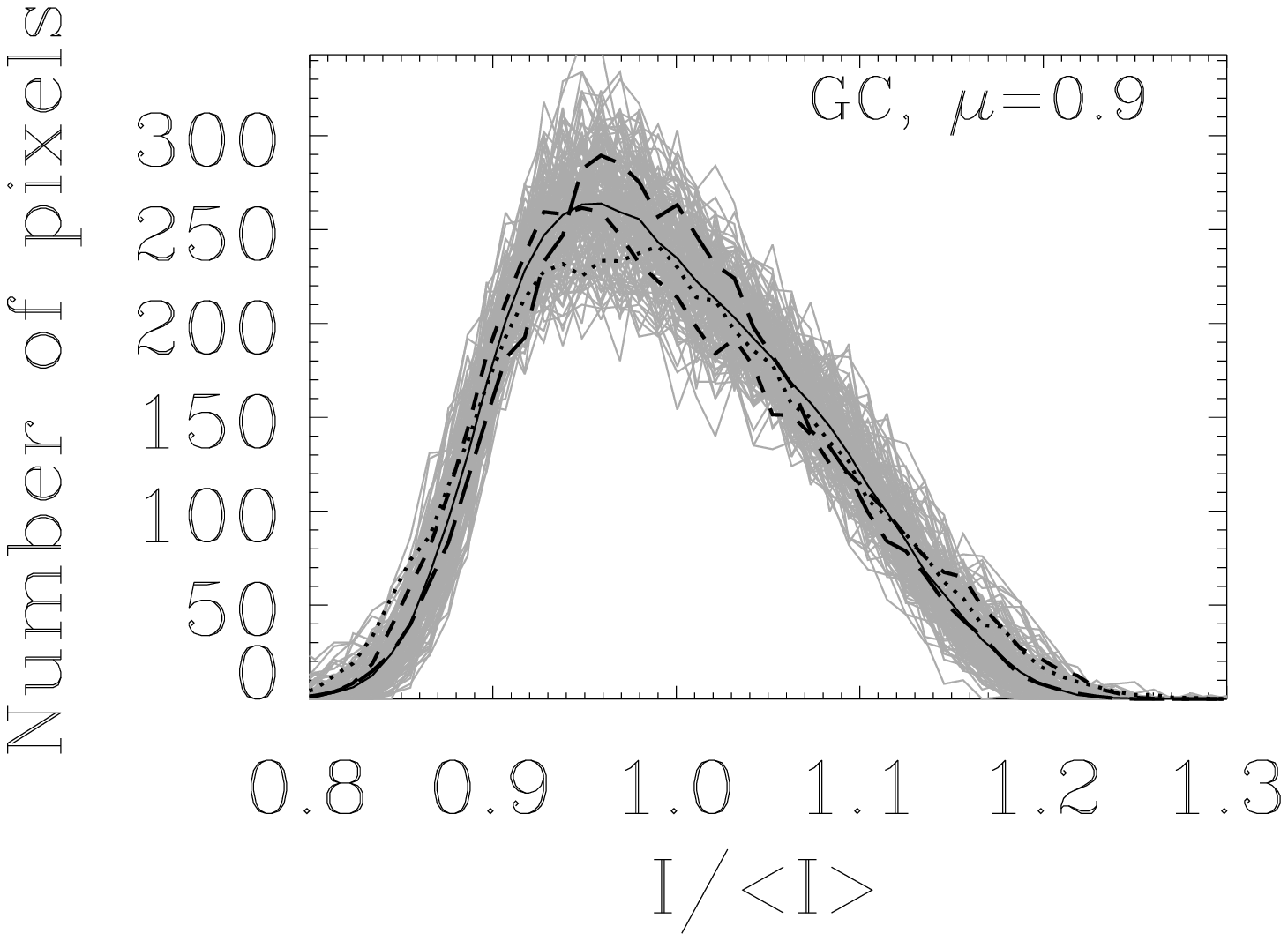} \end{picture}}
\put(-130,-110){\begin{picture}(0,0) \includegraphics{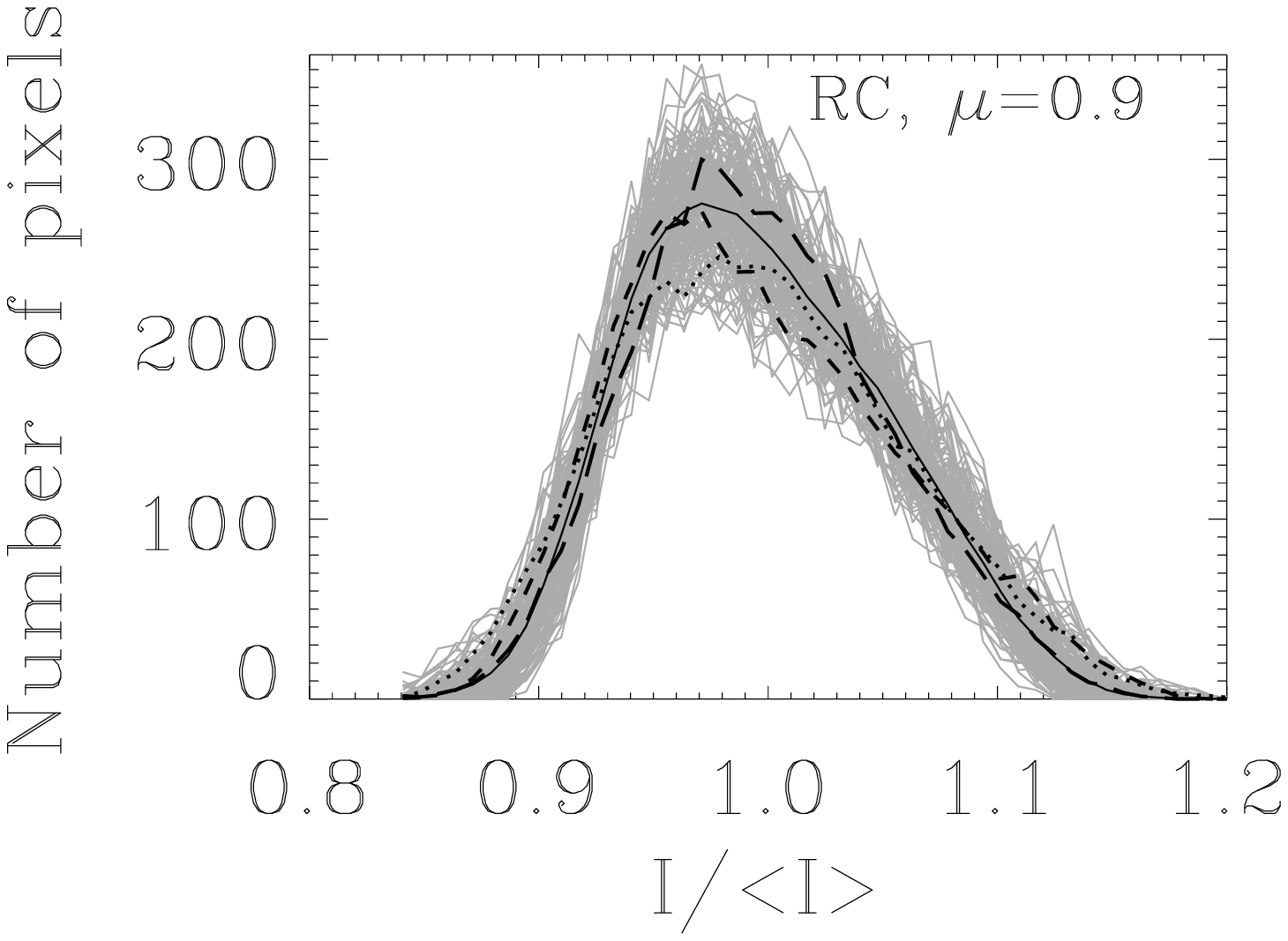} \end{picture}}
\put(20,290){\begin{picture}(0,0) \includegraphics{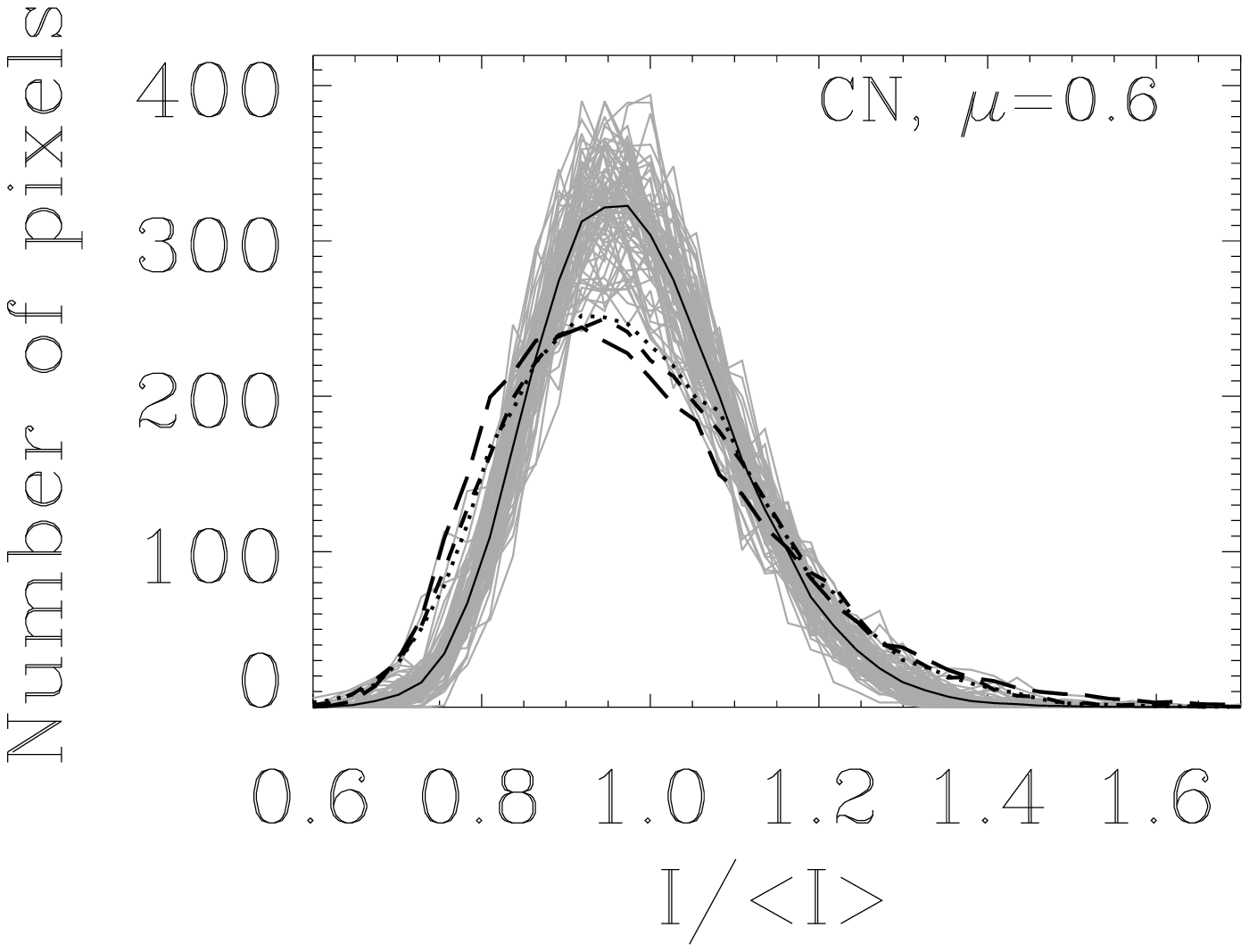} \end{picture}}
\put(20,190){\begin{picture}(0,0) \includegraphics{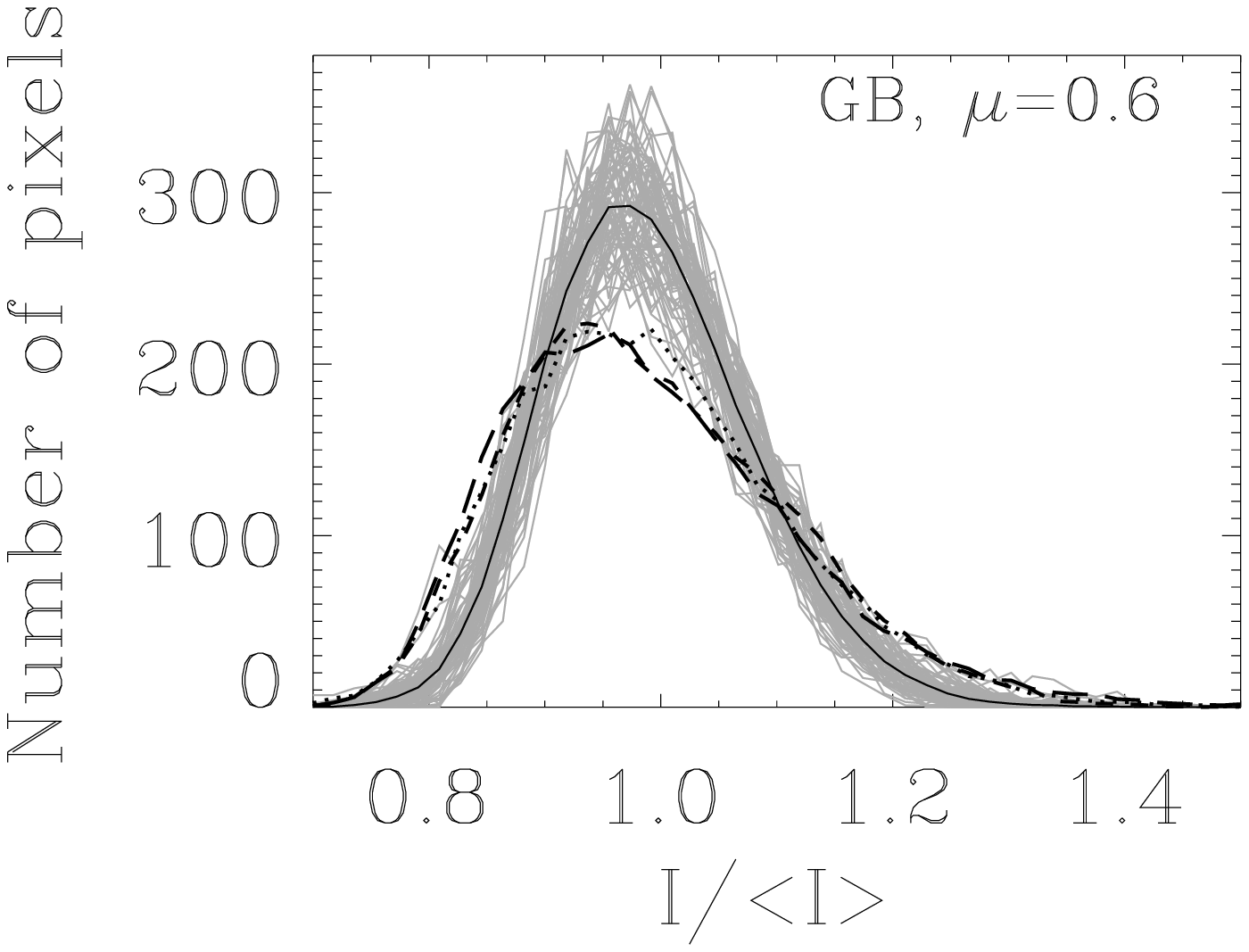} \end{picture}}
\put(20,90){\begin{picture}(0,0) \includegraphics{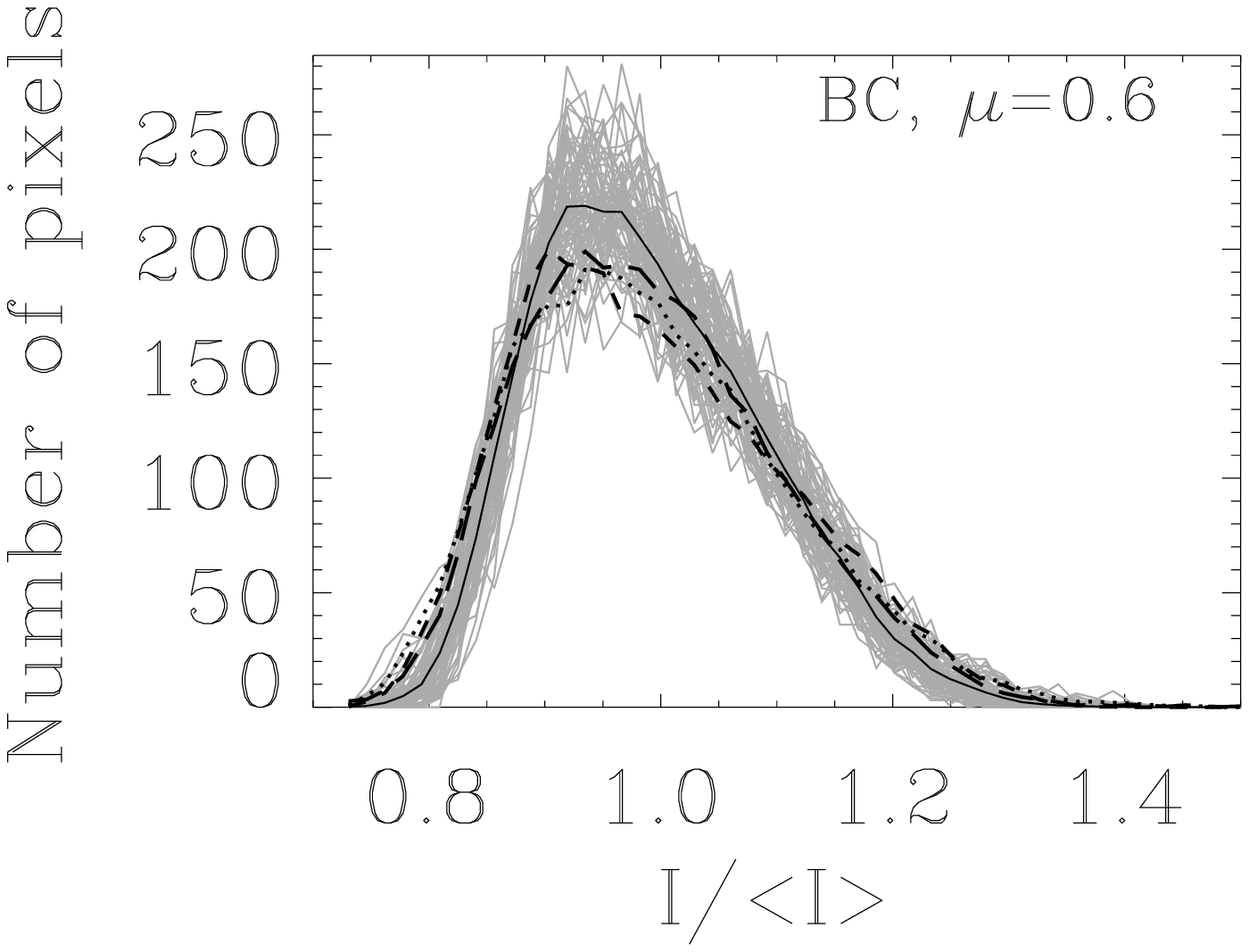} \end{picture}}
 \put(20,-10){\begin{picture}(0,0) \includegraphics{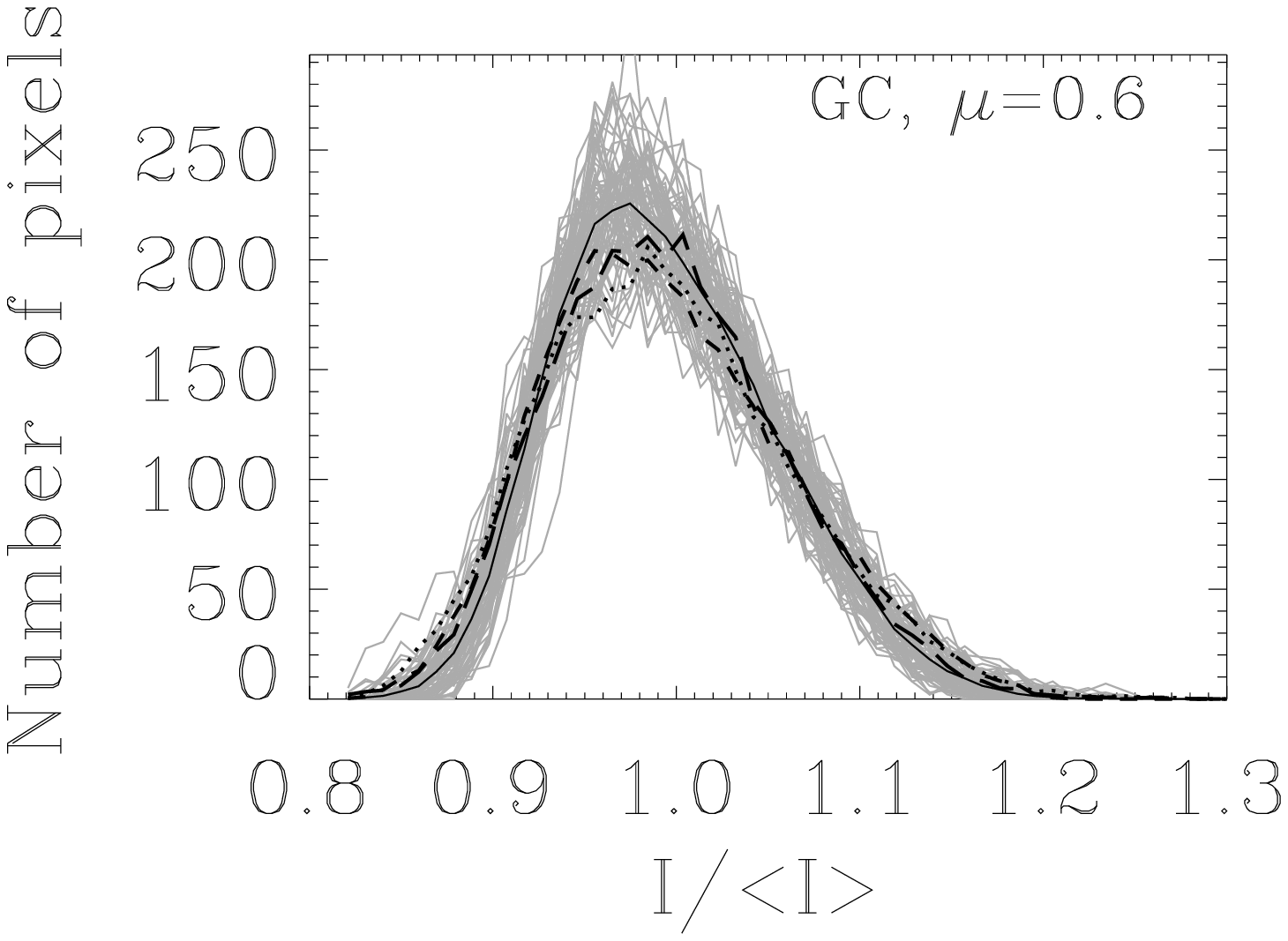} \end{picture}}
\put(20,-110){\begin{picture}(0,0) \includegraphics{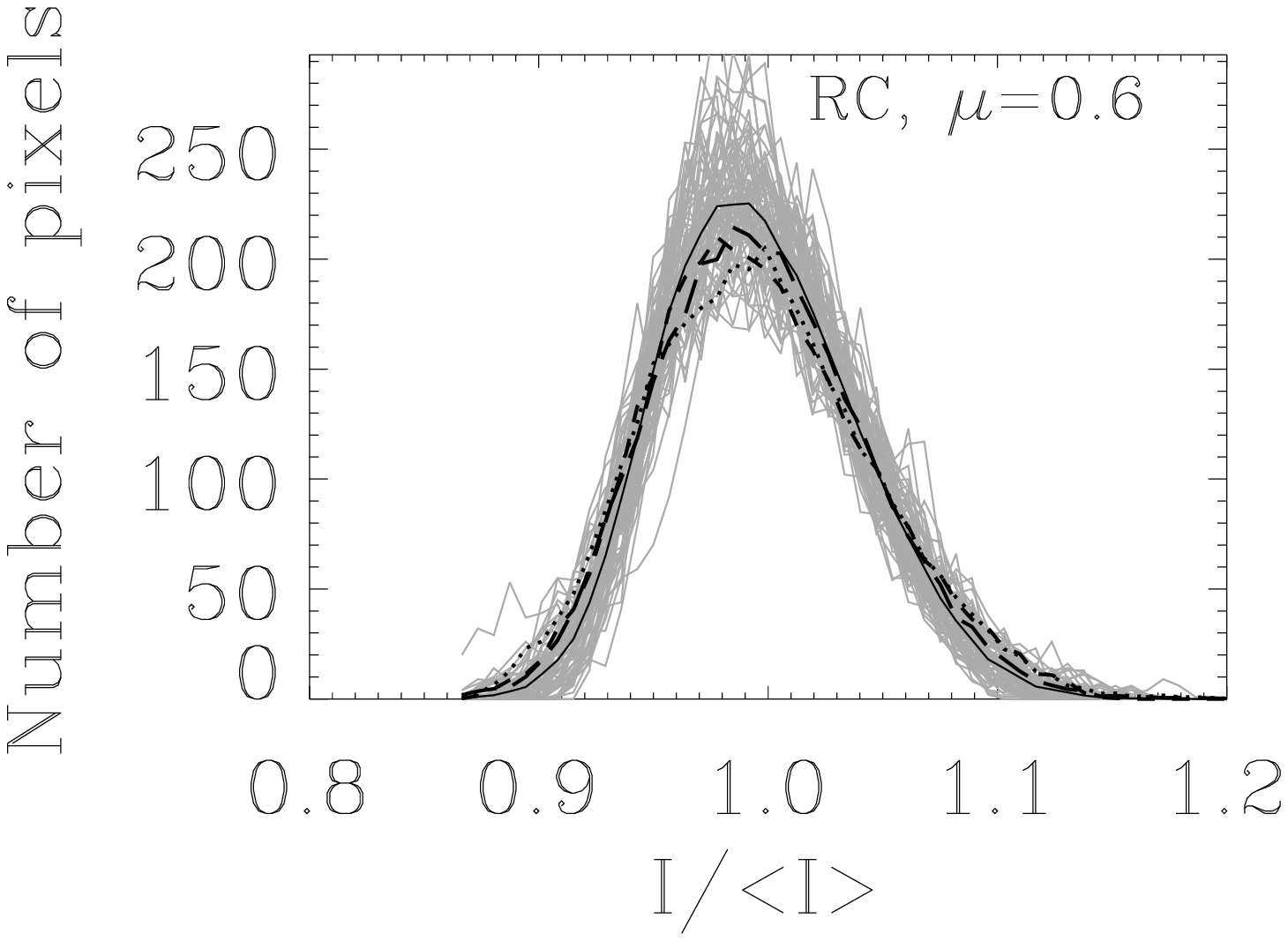} \end{picture}}
\put(170,290){\begin{picture}(0,0) \includegraphics{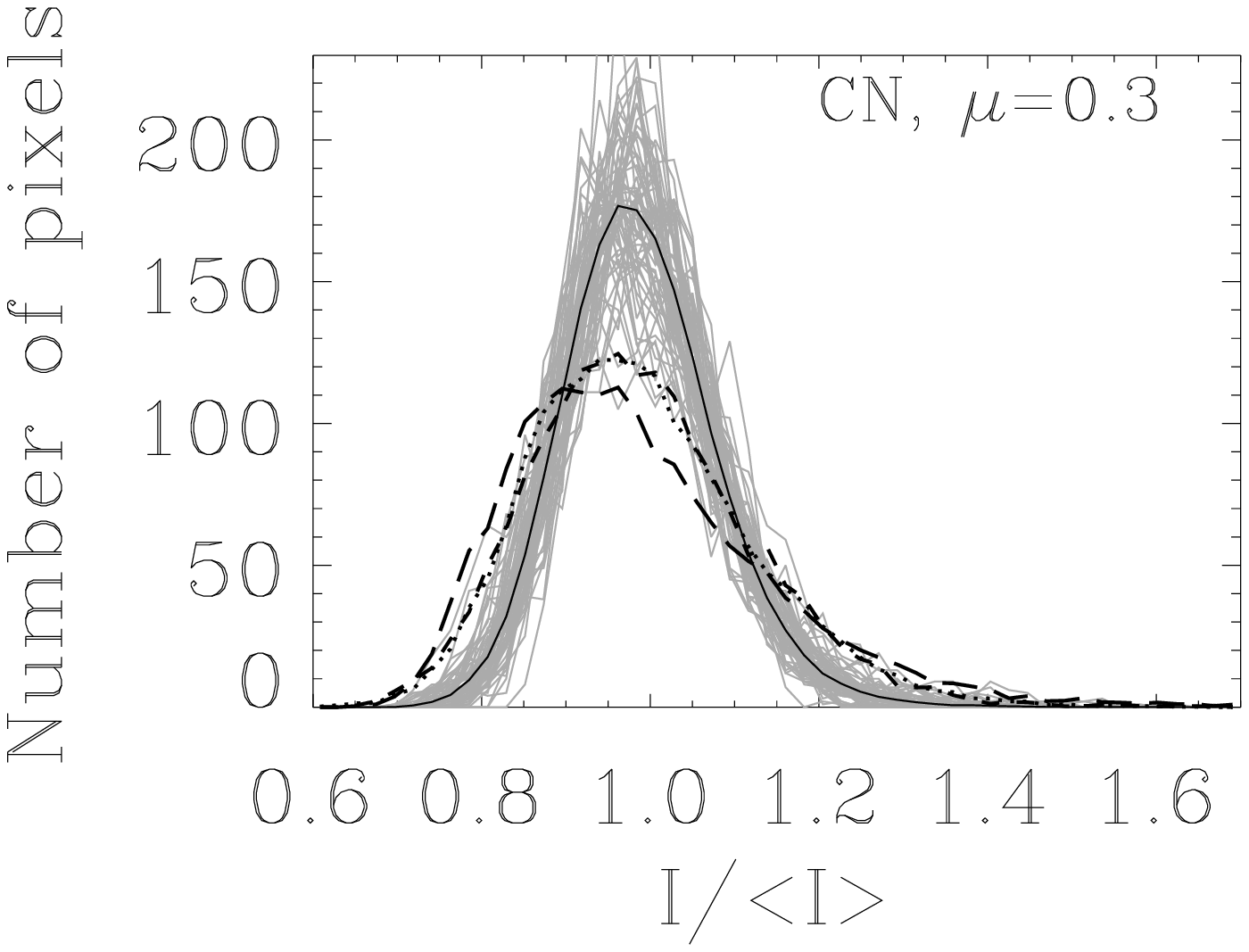} \end{picture}}
\put(170,190){\begin{picture}(0,0) \includegraphics{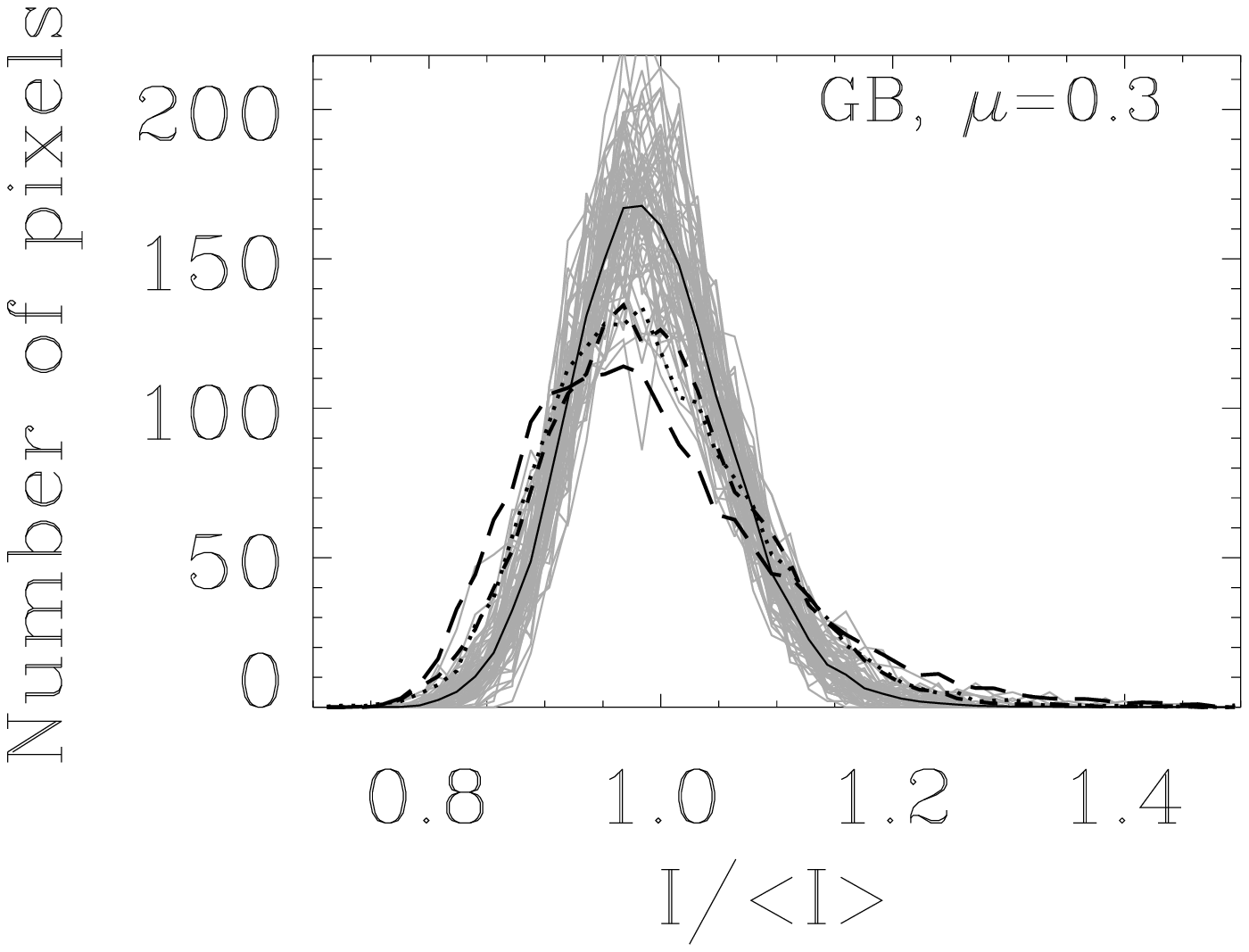} \end{picture}}
\put(170,90){\begin{picture}(0,0) \includegraphics{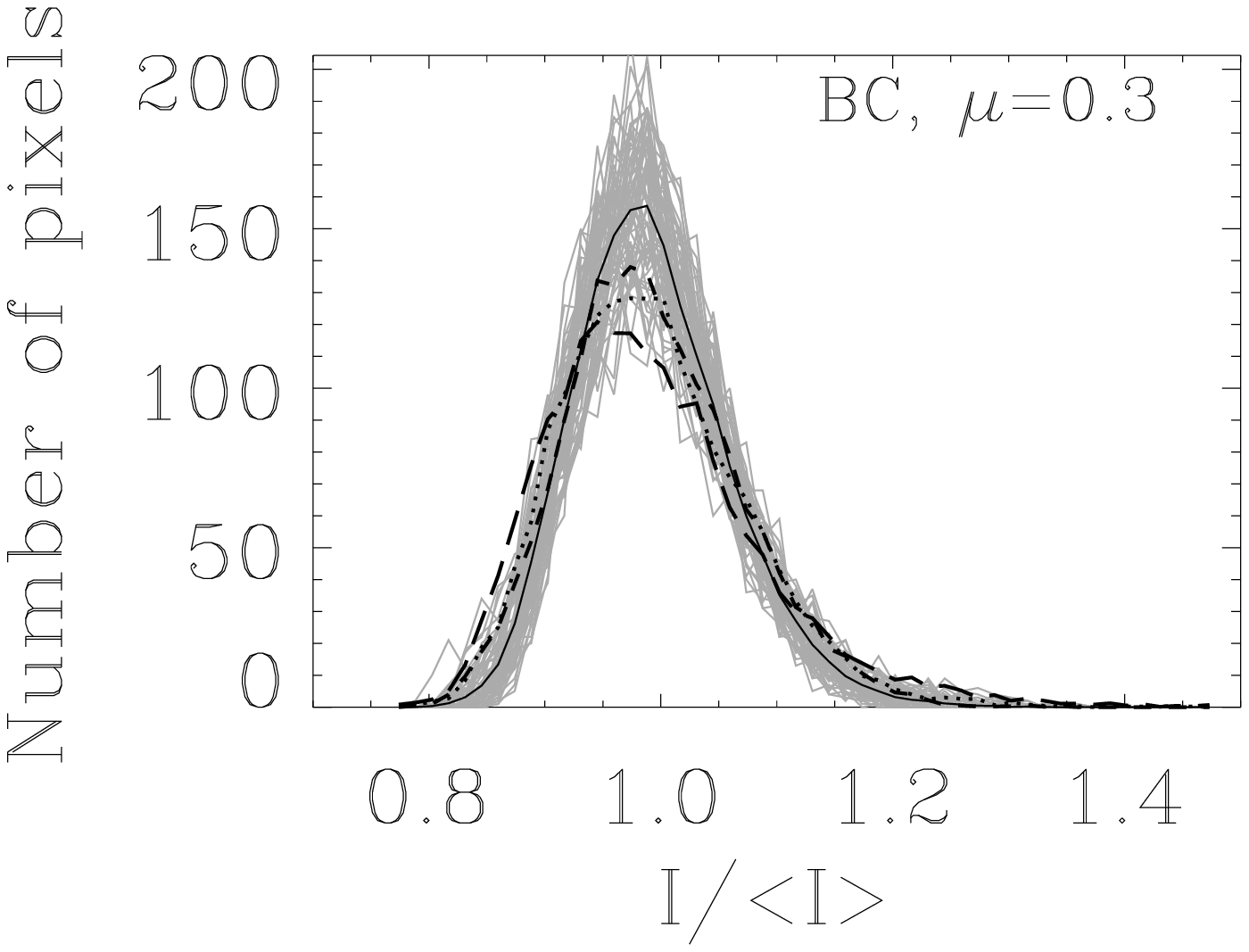} \end{picture}}
 \put(170,-10){\begin{picture}(0,0) \includegraphics{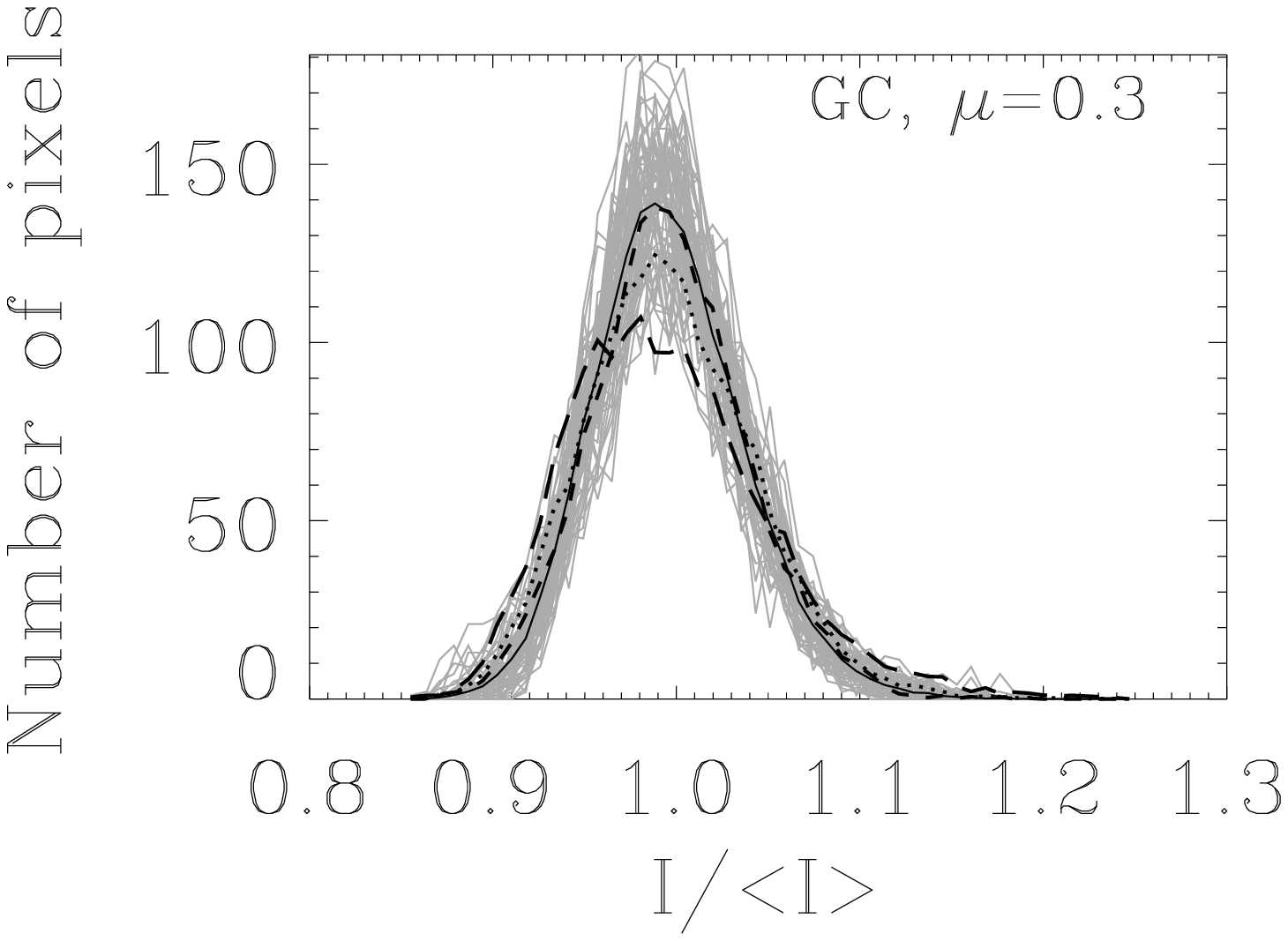} \end{picture}}
\put(170,-110){\begin{picture}(0,0) \includegraphics{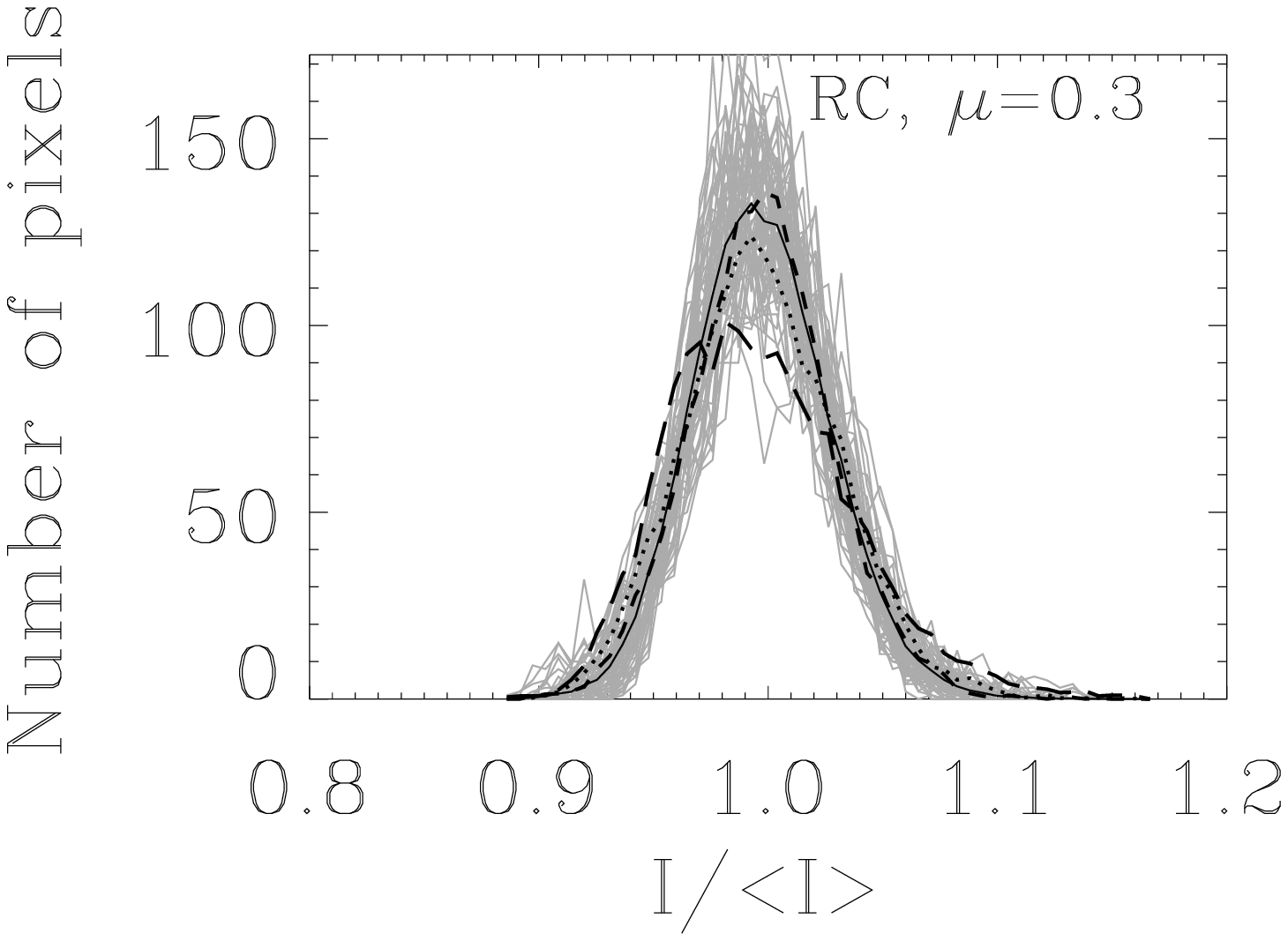} \end{picture}}
\end{picture}
\caption{Histograms of the intensity distribution in convolved simulations and in observations for the 5 Hinode/SOT wavelengths, from top to bottom: 388 nm, 430 nm, 450 nm, 555 nm, 668 nm. From left to right: Limb angles $\mu=0.9$, $\mu=0.6$, $\mu=0.3$. The histograms for the simulations are shown for 3 different magnetic fluxes: 0G (dotted), 50G (dashed), 200G (long dashes). To illustrate the spread in the observations, we overplot  histograms of extracts of the whole observed image (grey lines). The black solid line indicates the mean over all observed extracts. Note the different scales in the horizontal axes.}
\label{fig:muram_hinode_conv_histo_all}
\end{figure*}

\bigskip

\noindent \textbf{Intensities:} We calculate the emergent intensities along parallel rays, each passing through a different point of the surface grid of the MHD simulations. Since we are interested in a broad spectral coverage, we use Kurucz' spectral synthesis code ATLAS9 \citep{kurucz1993} (which includes diatomic molecules) with opacity distribution functions (ODFs) on each of those simulated model atmospheres at disk centre and towards the limb. ATLAS9 (as
rewritten by J. B. Lester) and the ODFs were obtained through
CCP7 (Collaborative Computing Project No. 7\footnote{http://ccp7.dur.ac.uk/}). The program solves the radiative transfer equations under the assumption of local thermodynamic equilibrium (LTE).  The simplification of LTE and the use of ODFs allows us to
calculate the whole spectrum from 160 nm to 160 000 nm with a
spectral resolution of better than 200 in the visible. However, the HINODE/SOT/BFI filters are narrower than the ATLAS9 resolution bins, and, in addition, the spectra are not calculated at the wavelength centre for the filters which leads to a comparison at slightly different wavelengths. Furthermore, the CN and GB regions comprise  many spectral lines that are only approximated by coarsely resolved ODFs in the ATLAS9 calculation. This may influence the comparison of the simulated intensities with the observations for the CN and GB wavelengths.

For an inclined view away from disk centre, we allow the line of sight to cut through the simulated atmosphere at an angle $\theta$ to the surface normal ($\theta$ is the heliocentric angle). For our simulation box, this means we have to interpolate the atmosphere parameters along rays with the respective heliocentric angle. 

For the comparison with observations, we degraded the simulated images to take into account optical effects: The point-spread functions for the SOT as calculated by \cite{mathewetal2009} were used for both dataset I and dataset II, as discussed in Sect.~2.

%%%%%%%%%%%%%%%%%%%%%%%%%%%%%%%%%%%%%%%%%%%%%%%%%%%%%%%%%%%%%%%%%
    
\section{Comparisons of observations with simulations}
%%%%%%%%%%%%%%%%%%%%%%%%%%%%%%%%%%%%%%%%%%%%%%%%%%%%%%%%%%%%%%%%%

%\begin{figure*}
%\centering
%\begin{picture}(200,200)
%\put(-190,-180){\begin{picture}(0,0) %\special{psfile=clv_50G200G400G_ermolli_bluePSPT_0.1-1.0_bw_legend.eps
%     hscale=50 vscale=50 voffset=0} \end{picture}}
%\put(50,-180){\begin{picture}(0,0) %\special{psfile=clv_50G200G400G_ermolli_redPSPT_0.1-1.0_bw_legend.eps
%     hscale=50 vscale=50 voffset=0} \end{picture}}
%\end{picture}
%\caption{Centre to limb variations for the blue and red continuum wavelengths, respectively, for average vertical magnetic fields of 50G, 200G, and 400G. Overplotted are the data analysed in \cite{ermollietal2007}. The contrast increases towards the limb and is higher for shorter wavelengths.}
%\label{fig:clv_50G_200G_400G_ermolli}
%\end{figure*}  

Here, we compare the resulting intensities from simulations with observations to validate the simulated results. For dataset I at $\mu=0.9$ with its finer binning, we first inspect the observed and simulated intensity images visually for a qualitative comparison. In a next step we compare dataset II with the simulations (at all limb angles) in terms of rms contrasts and intensity distributions. 

\subsection{Simulations vs. dataset I: Visual comparison}
\noindent In this section, we compare observations with original and convolved simulations at a limb angle of $\mu=0.9$.  In Fig.~\ref{fig:muram50G_hinode_unconv_conv_all} original and convolved \textit{MURaM} simulation snapshots for 50G  are shown together with the observed image in the respective wavelength filter. The qualitative agreement is good, the granules are of comparable size and the visually perceived contrasts appear to be similar. The occurrence of brighter and darker features is similar and also changes correspondingly for the different wavelength filters. The brightest features can be seen in the CN wavelength. The bright magnetic features that appear for the 50G (as seen in Fig.~\ref{fig:muram50G_hinode_unconv_conv_all}) and 200G simulations are not seen for nonmagnetic simulations.

\subsection{Simulations vs. dataset II: rms contrasts and intensity distributions at different limb angles}

In Fig.~\ref{fig:hist_gc_unconv} we demonstrate the effect of convolving with the PSF in a more quantitative way. We plot, as an example, the histograms for the green continuum intensity distribution in the original simulation  at the  limb angle   $\mu=0.9$ for  0G, 50G, and 200G. We overplot (in grey) the corresponding histogram for the convolved simulation as shown in Figure \ref{fig:muram_hinode_conv_histo_all}.  As expected, the convolved histograms are significantly narrower. For most wavelengths we find a distribution with a double peak in the unconvolved simulations, which is most prominent in the 50G simulation, and represents the dark intergranular components (the more prominent peak at $I/\langle I \rangle $ $<$1) and the bright granular components (the weaker peak at $I/\langle I \rangle $ $>$1) (see e.g. \cite{wedemeyervoort2009} and references therein).  After the convolution, the double-peak is smeared out, but important differences between the nonmagnetic, 50G, and 200G simulations remain, in particular for the visible broadband distributions (BC, GC, and RC). 

Figure ~\ref{fig:muram_hinode_conv_histo_all} shows the histograms of the intensity for (all) 10 averaged simulation snapshots and for average magnetic fields of 0G, 50G, and 200G. From left to right we plot histograms for the convolved simulations at the three limb angles $\mu=0.9$, $\mu=0.6$, $\mu=0.3$ (i.e. from disk centre towards the limb) and for all 5 Hinode/SOT wavelengths (from top to bottom: CN: 388 nm, GB: 430 nm, BC: 450 nm, GC: 555 nm, RC: 668 nm). We also show the averaged histograms for the Hinode observations (dataset II, solid black line).  We illustrate the spread in the observations by  overplotting  histograms of extracts of the observed image (grey lines). For ease of comparison, the simulations have been renormalised to cover the same number of points as the observations.  

 \begin{figure}
\centering
\begin{picture}(200,170)
\put(-70,-180){\begin{picture}(0,0) \includegraphics{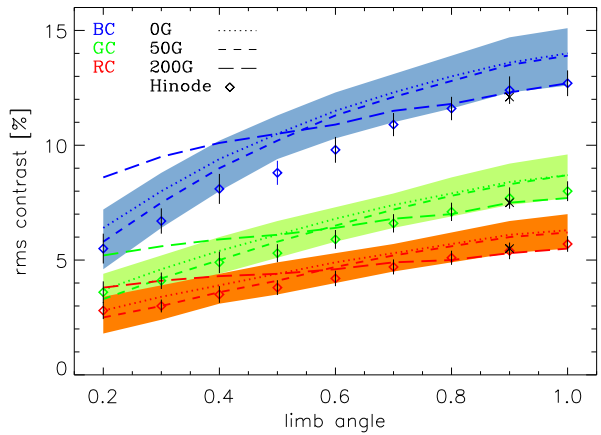} \end{picture}}
\end{picture}
\caption{Rms contrasts for the BC, GC, and RC  Hinode/SOT wavelengths at the limb angles  $\mu=0.2-1.0$ for the mean over the 0G simulation snapshots (dotted lines),  the 50G simulation snapshots (dashed line),  the 200G simulation snapshots (long dashes) and the Hinode dataset II observations (diamonds). We show the spread of the 10 snapshots of the 50G simulation and overplot the 1$\sigma$ error of the Hinode dataset II observations. Overplotted (black crosses) are the Hinode dataset I results at $\mu=0.9$, which are almost indistinguishable from the dataset II results.}
\label{fig:rms_all_limbangles}
\end{figure}

 \begin{figure*}
\centering
\begin{picture}(200,90)
\put(-180,-120){\begin{picture}(0,0) \includegraphics{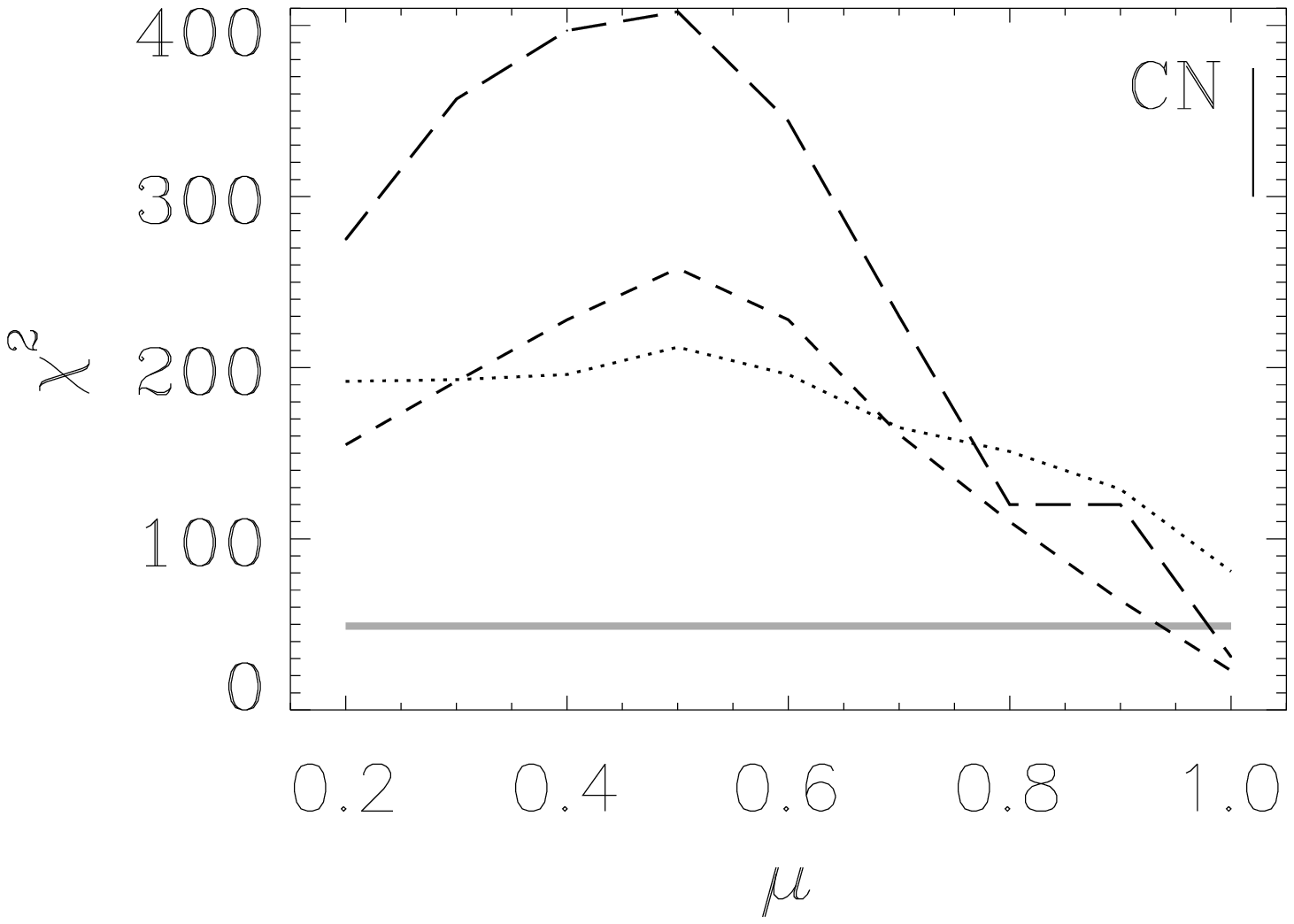} \end{picture}}
\put(-80,-120){\begin{picture}(0,0) \includegraphics{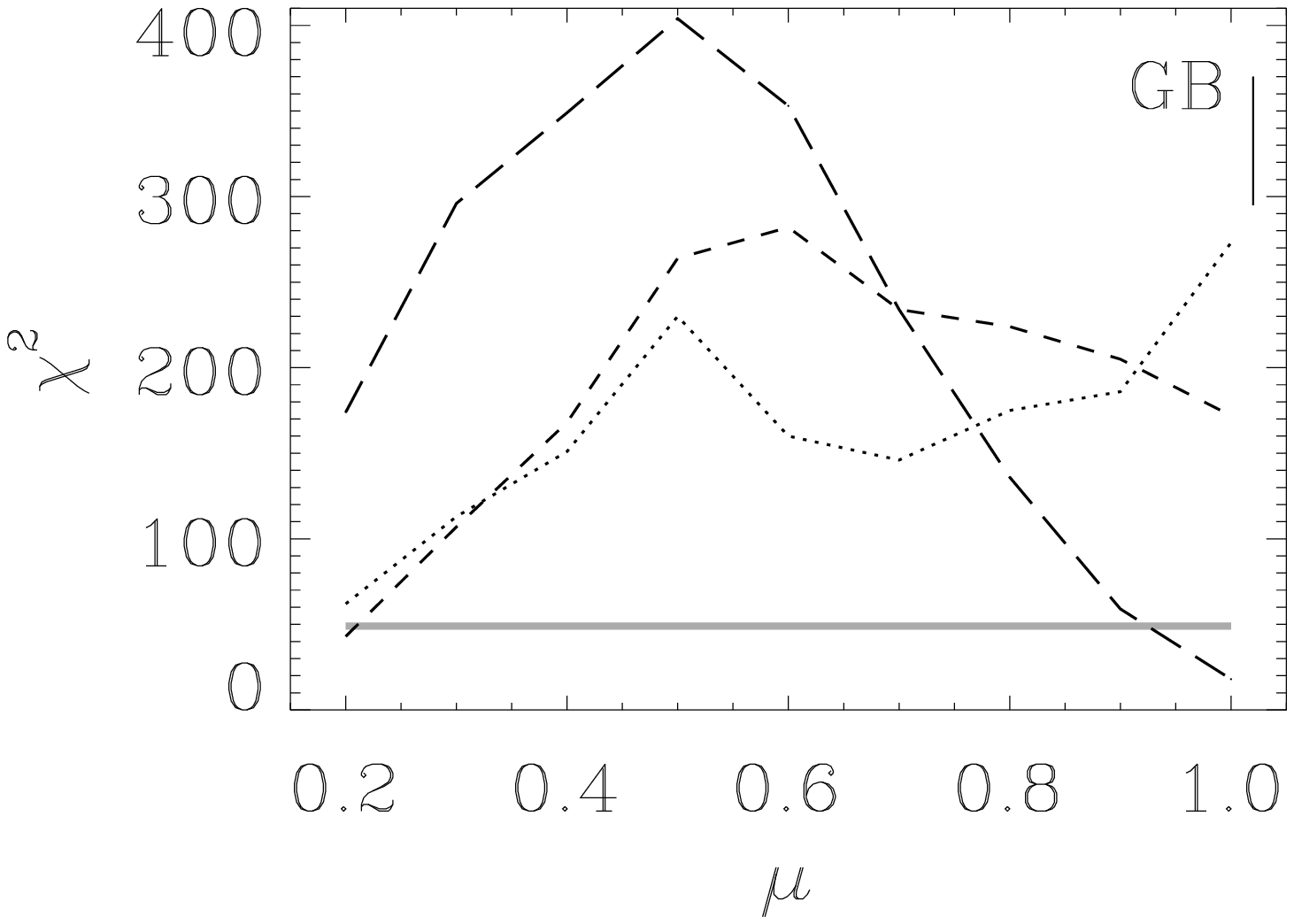} \end{picture}}
\put(20,-120){\begin{picture}(0,0) \includegraphics{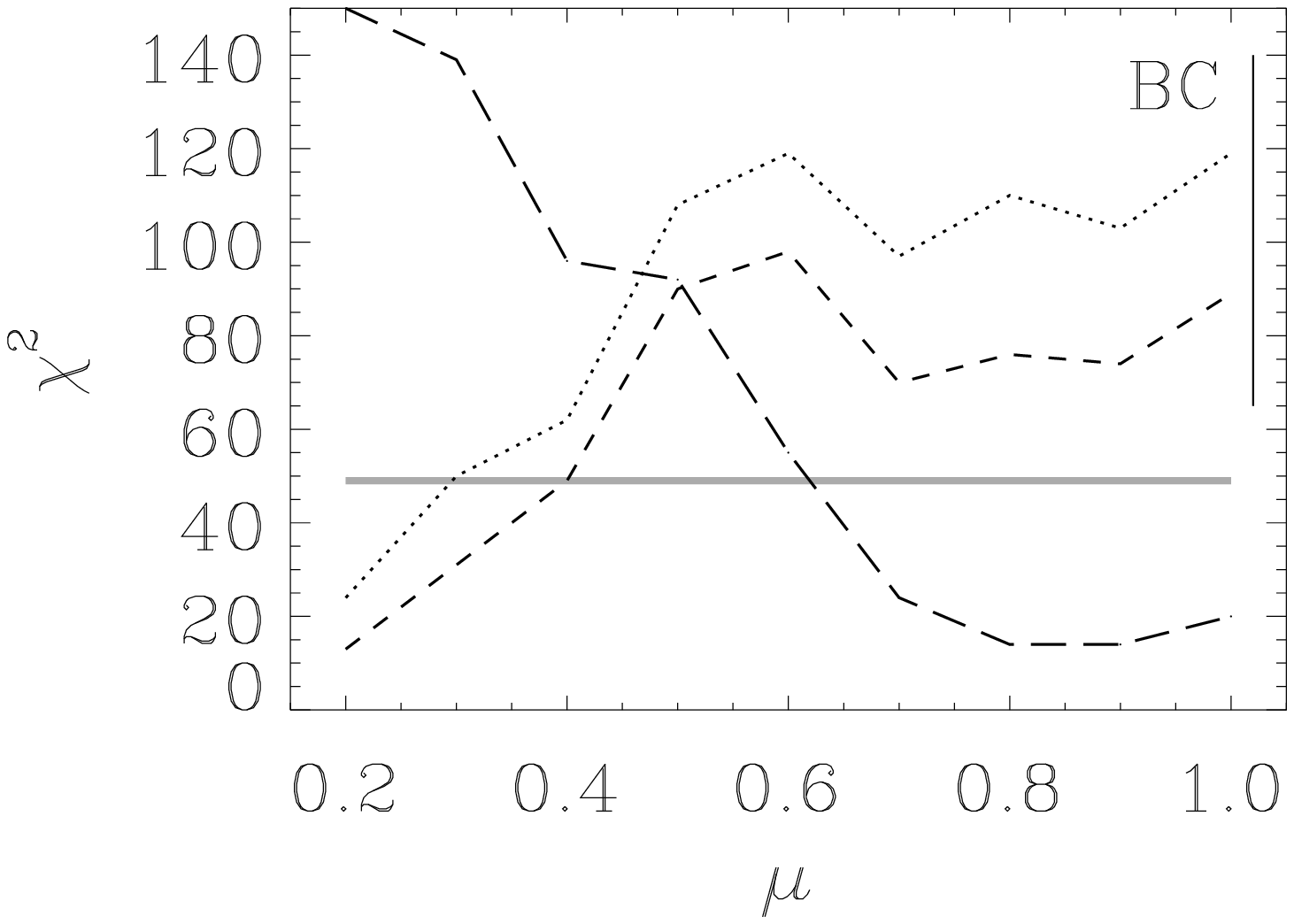} \end{picture}}
\put(120,-120){\begin{picture}(0,0) \includegraphics{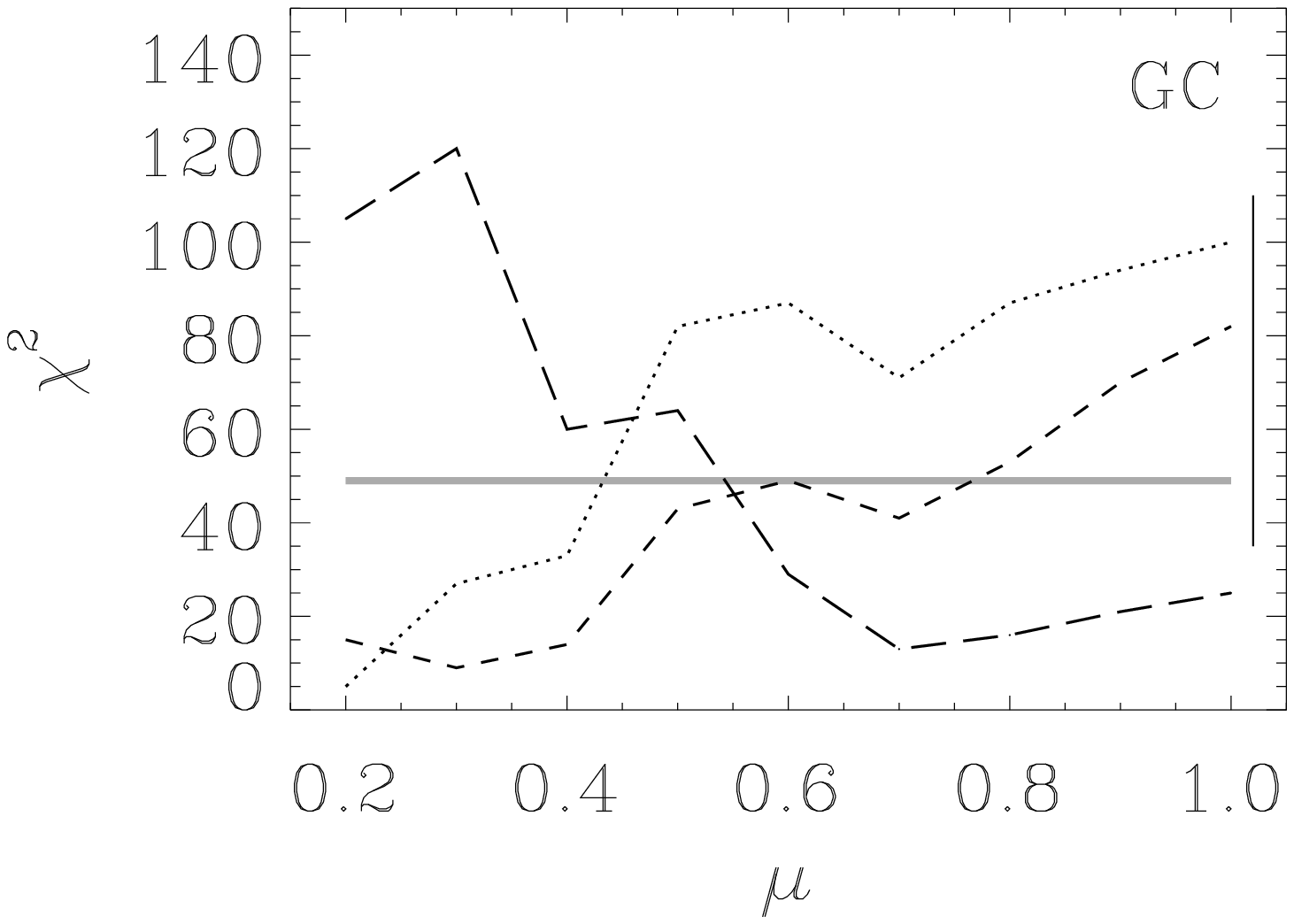} \end{picture}}
\put(220,-120){\begin{picture}(0,0) \includegraphics{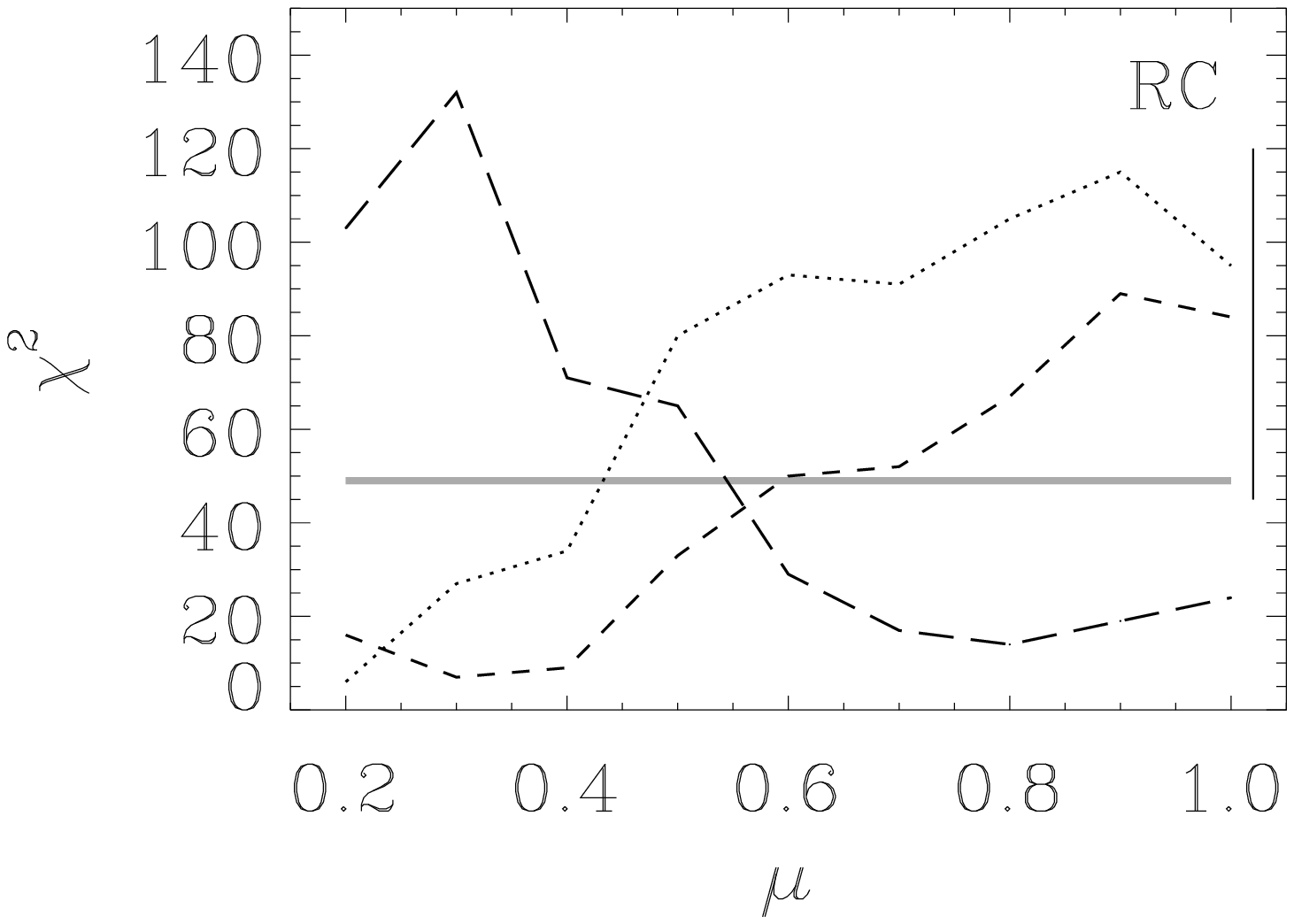} \end{picture}}
\end{picture}
\caption{$\chi^2$ values for  histograms of convolved simulations as shown in Fig.~\ref{fig:muram_hinode_conv_histo_all} but for all the limb angles  $\mu=0.2-1.0$. The $\chi^2$ values for 0G, 50G, and 200G are represented by dotted, dashed, and long-dashed lines, respectively. The $\Delta$$\chi^2$ for a confidence level of 99\% is 75 for 50 bins. A $\Delta$$\chi^2$ of 75 is indicated by the solid line on the right of each plot. The grey line at $\chi^2=49$ indicates the level where $\chi^2_{\rm r}=1$.}
\label{fig:chi}
\end{figure*}

At the limb angle of $\mu=0.9$, the agreement between data and simulations is reasonable for all wavelengths. Closer to the limb, at  $\mu=0.6$ and $\mu=0.3$, the histograms for the simulations in the CN and GB wavelengths do not agree with the measurements. For the three continuum filters, however, the agreement is satisfactory, as the histograms for the simulations all fall within the scatter of the data.  None of the simulations exactly matches the observed histograms,  which leads to high $\chi^2$ values.  A quantitative comparison and the $\chi^2$ analysis are discussed in the next section.

Intensity variations between bright and dark features determine the root-mean-square (rms) contrast, which is the standard deviation divided by the mean value. Figure \ref{fig:rms_all_limbangles} shows the rms contrasts for the convolved 0G, 50G, and 200G simulations in comparison to the resulting rms contrast from the observations at the limb angles $\mu=0.2-1.0$. For the simulations, these rms contrasts are the averages over rms contrasts of all simulation runs. For the observations, we calculate the rms contrasts for each of the extracts of the observed images, i.e. extracts of similar size as one simulation snapshot, and then average them.  

 \begin{figure*}
\centering
\begin{picture}(200,200)
\put(-190,-180){\begin{picture}(0,0) \includegraphics{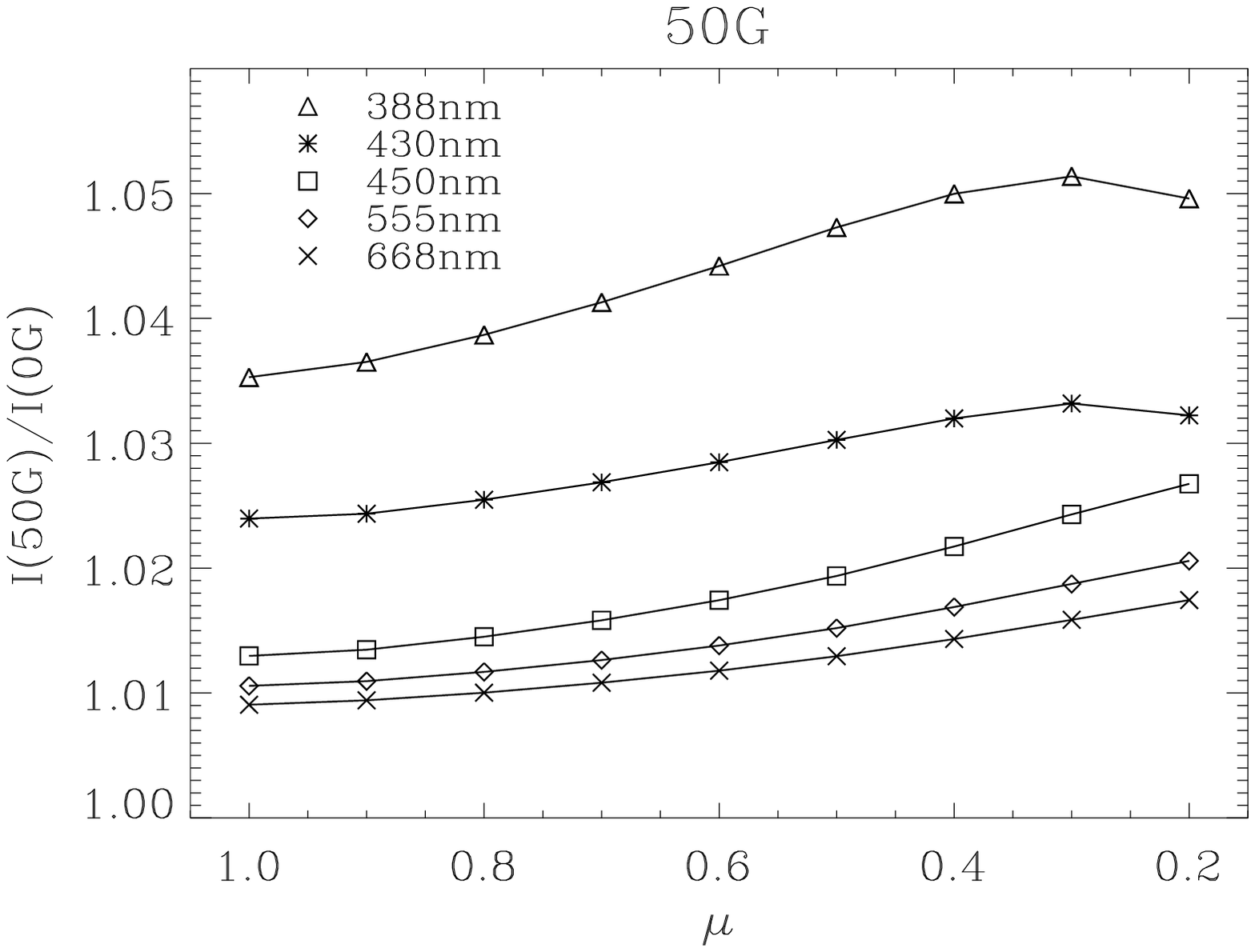} \end{picture}}
\put(50,-180){\begin{picture}(0,0) \includegraphics{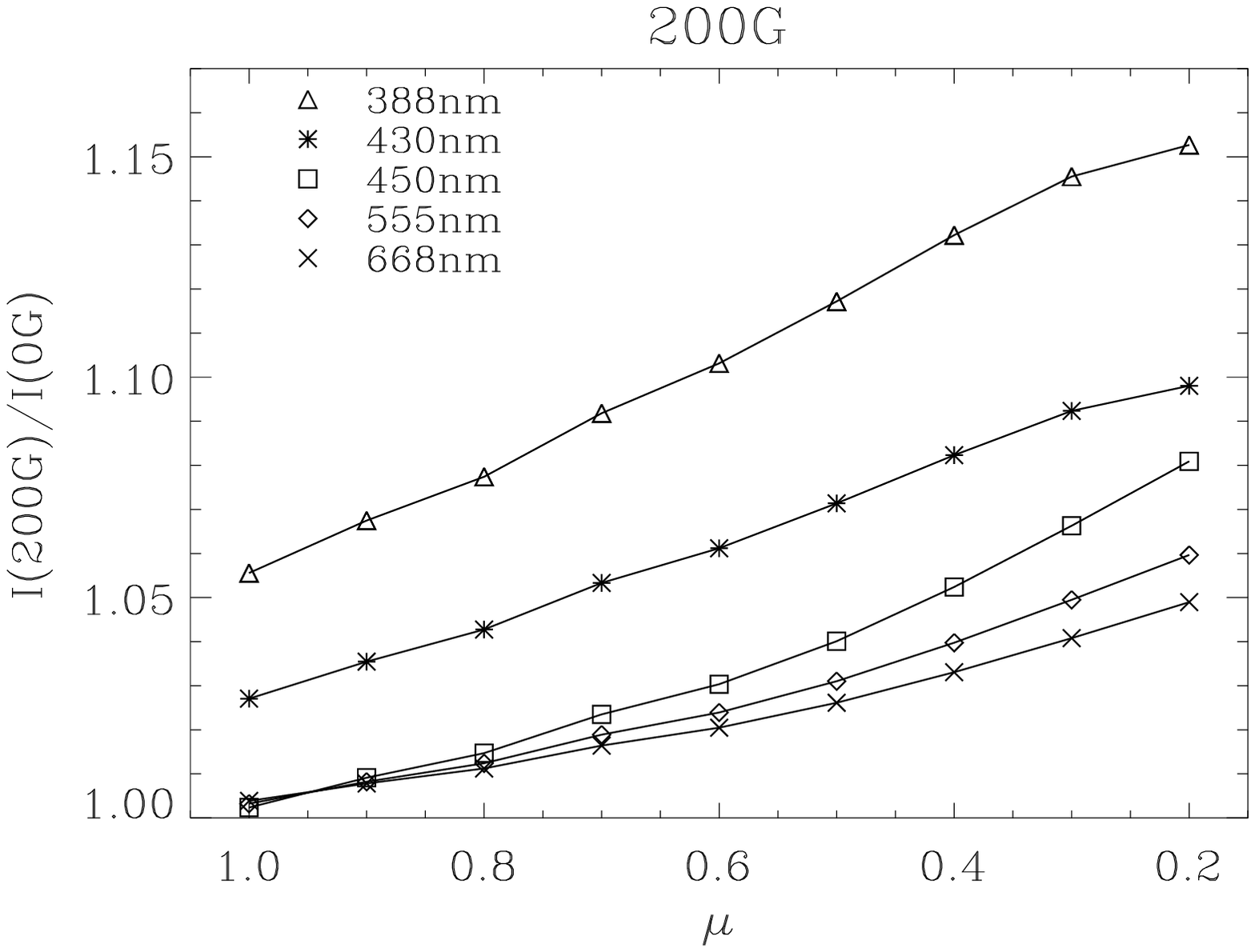} \end{picture}}
\end{picture}
\caption{Centre to limb variations for the five different Hinode/SOT wavelengths for an average vertical unmagnetic field of 50G and 200G, respectively. Note the different scales of the y-axes.}
\label{fig:clv_50G_200G}
\end{figure*}

 We consider the change in the rms contrast as one moves from weaker to higher magnetic fluxes, as is indicated by the dotted, dashed, and long-dashed lines in Fig.~\ref{fig:rms_all_limbangles} for the 0G, 50G, and 200G simulations, respectively. It can be seen that the 200G rms contrast is not a simple extrapolation of the 0G and the 50G contrasts. The 0G and  50G rms contrasts look very similar, although the 50G rms contrast falls off a bit more steeply. The 200G rms contrast shows a much flatter profile with a lower contrast at disk centre and a higher contrast at the limb compared to the weaker fields. The  rms contrasts suggest that the 50G simulation agrees slightly better with the Hinode observations than the 0G simulation for all wavelengths. Although the 200G simulation fits the observations better near disk centre (as can also be seen from the  $\chi^2$ statistic in the next section), its overall behaviour does not represent the one seen in the data.  Given that the observations refer to quiet Sun, this result is not surprising. The agreement between the contrast as measured from the observations and the simulation for 50G is good in the visible continuum bands at 450.4 nm (BC), 555.0 nm (GC), and 668.4 nm (RC). The contrasts derived  for 388.3 nm (CN) and for 430.5 nm (GB) are generally too large. For the 50G simulation, they can exceed the measured contrasts by up to a factor of 1.5 and 1.3 for CN and GB, respectively, depending on limb angle.

\subsection{Statistics for simulations}

Although the differences between the simulations with different average 
magnetic fluxes are less distinct after the convolution, they can still be 
seen in the histograms in Fig.~\ref{fig:muram_hinode_conv_histo_all}. To 
find the simulations that agree best with the observations, we calculate 
the value of $\chi^2$ according to Eq.~1, where $R_i$ and $S_i$ are the number of pixels in intensity bin $i$ for the observations and simulations, respectively.
The $\chi^2$ values are shown in Fig. \ref{fig:chi} for limb angles of 
$\mu=0.2-1.0$ and for the five Hinode wavelength filters.  
%[AAA]
As each histogram has 50 bins, i.e., 49 degrees of freedom (d.o.f), 
we would expect a good fit to have a value of $\chi^2 \simeq 49$. This 
corresponds to a value of approximately one for the reduced $\chi^2$ given 
by $\chi^2_{\rm r} = \chi^2 / \rm{d.o.f}$. 
%[DO WE USE THE REDUCED 
%$\chi^2$? WE MIGHT NOT NEED TO? IF YOU DECIDE TO GET RID OF $\chi^2_r$ 
%CHANGE THE TEXT BETWEEN THE AAAs TO WHAT IS ATTACHED AT THE VERY END 
%OF THIS MAIL. ALSO, PLEASE CHECK WHETHER THAT d.o.f ABBREVIATION LOOKS 
%OK.. MIGHT BE BETTER TO USE dof].
The grey line in Fig. \ref{fig:chi} indicates the level where $\chi^2_{\rm 
r}=1$.
% [AAA]

It turns out that, except at disk centre, none of the simulations fit the 
CN or G-band observations well. This is not too surprising as we are 
currently assuming LTE in our spectral synthesis calculations, and are, 
furthermore, not carrying out a full line-by-line synthesis, but use ODFs 
to approximate the opacity in relatively coarse wavelength bins (see 
Sect.~3). For the blue, green and red continuum 
bands, the simulations agree reasonably well, though it appears that the 
200G simulation fits best for positions closer to disk centre, while the 
0G and 50G simulations show better agreement closer to the limb.

We can use contours of equal $\chi^2$ to examine whether the convolved 
intensity distributions resulting from different magnetic flux levels are 
distinguishable, i.e., appear to be drawn from a different underlying 
population. To this aim, we calculate $\Delta \chi^2 = \chi^2 - \chi^2_m$, 
where $\chi^2_m$ denotes the minimum $\chi^2$. In our case of 50 bins, 
$\Delta \chi^2$ values in excess of 75 suggest a different distribution at 
a confidence level of 99\% \citep[see][]{pressetal}. 
The 99\% confidence levels are indicated by the vertical bars on 
Fig.~\ref{fig:chi}. They confirm that the intensities calculated for the 
CN filter are not a good representation of the observations, except near 
disk centre, where the convolved non-magnetic, 50G and 200G 
distributions are indistinguishable at the 99\% confidence level. In all 
other bands, the non-magnetic and the 50G convolved distributions are 
largely indistinguishable, while the 200G distributions appear to show 
significantly different behaviour, in particular towards the limb (see 
also the right-hand column in Fig.~\ref{fig:muram_hinode_conv_histo_all}). While the 50G simulation does 
not represent the best fit at most limb angles, it largely lies within 
$\Delta \chi^2 =75$. 
%-----------------------------------------------------------------
%ALTERNATIVE BIT WITHOUT REDUCED CHI-SQ FOR PASSAGE BETWEEN [AAA]s:
%As each histogram has 50 bins, i.e., 49 degrees of freedom, we
%would expect a good fit to have a value of $\chi^2 \simeq 49$, as 
%indicated by the grey lines in Fig.~\ref{fig:chi}.
%NOTE THAT YOU SHOUD THEN ALSO CHANGE THE LAST SENTENCE IN THE CAPTION 
%OF FIG 7.

%The number of data points for the simulations is $\mu$  $\times $ 288 $\times$ 288, and $\mu$  $\times $ 76 $\times$ 76 for the observations, distributed over 50 bins. 
% \footnote{http://www.nr.com/} 

%even though the overall histogram for the RC looks more similar to the 50G simulations, the deviations between the 50G simulations and the observations in the wing for the very bright features with $I/\langle I \rangle $ $<$1.1 are quite large, resulting in an overall better $\chi^2$ value for the 200G simulations.

 \section{Intensity contrasts in simulations}

The centre-to-limb variation of the emergent intensity varies with wavelength. In the UV,  magnetic features are  brighter than at other wavelengths. These differences become even stronger near the limb and are therefore crucial for spectral solar irradiance variations. It is generally not possible to reproduce the observed limb behaviour with 1-D model calculations, as they tend to overestimate the limb contrast at extreme limb angles.

In Figure \ref{fig:clv_50G_200G} we show the simulated contrasts (i.e. the mean intensity of the 10 averaged magnetic 3-D MHD simulation snapshots compared to the mean quiet Sun ($=0G$) intensity) with average vertical magnetic field of 50G and 200G as a function of heliocentric position in the five Hinode wavelength filters. As expected, the contrast increases towards the limb and is higher for shorter wavelengths. The result is qualitatively consistent with figure 4 in \cite{voegleretal2005ii} if one assumes that the bolometric intensity corresponds roughly to the maximum of the Planck curve. Contrary to 1-D models, the flattening and turnover (e.g. for 50G at 388 nm and 430 nm) are in better agreement with measurements of facular contrasts than 1-D calculations \citep{topkaetal1997, unruhetal1999, ortizetal2002}.
%In Figure \ref{fig:clv_50G_200G_400G_ermolli},  we plot the simulated contrasts  as a function of heliocentric position in two different wavelengths, corresponding to the red and blue continuum PSPT (Precision Solar Photometric Telescope) filter \citep[see][]{ermollietal2007}. In the same figure, we overplot the data analysed in \cite{ermollietal2007} which mainly lie between the limiting contrasts resulting from the 50G and 200G simulations. The different linestyles symbolise the five different automated methods utilised to identify facular regions in pre-processed images as described in \cite{ermollietal2007}).
 %For the simulation with the average vertical magnetic field of 400G, the contrasts are significantly larger and do not follow the observations. The similar limb behaviour in observations and simulations shows that the approach via 3-D MHD simulations is more realistic than previous 1-D studies.   

%\begin{figure}[!h]
%\centre+ing
%\resizebox{\hsize}{!}{\includegraphics{1sig3sig.eps}}
%\caption{Example for one of the 50G snapshots (the one closest to the averaged 10 snapshots). Left: original, middle: the pixels outside the 3 sigma limits are black, right: the pixels outside the 1 sigma limits are black.}
%\label{fig:1sig3sig}
%\end{figure}

\section{Conclusions and Outlook}

In the present paper, we have compared simulated and observed intensity distributions in the quiet Sun at a range of limb angles and derived mean contrast for simulations with an average magnetic field of 0G, 50G, and 200G.  

 Comparing histograms and rms contrasts of the observations and simulations, we find that no single simulation reproduces the data for all limb distances and colour bands. The 50G simulation intensity histograms and rms contrasts agree best with the observations near the limb and show an overall similar behaviour. The 200G simulation, however,  reproduce the observations better near disk centre, but do not represent the overall behaviour of the observed rms contrasts. 

There may be various reasons for the discrepancy between the data and the simulations. One of these could  be the use of LTE in ATLAS9 along with the fact that we are not calculating spectral lines (which mainly affects the results for CN and GB). Also, in our dataset we are observing different areas on the Sun at various limb angles with potentially differing activity levels. In additon, by averaging over a large area, the Hinode histograms represent a de facto mixture of features/magnetic fluxes, that we would expect to be reproduced by a combination of the simulations with 0G/50G/200G average magnetic fields. Furthermore, the height of the model atmosphere in our simulation box might be insufficient. As the radiation is produced at increasingly higher layers in the atmosphere as one moves to the solar limb and to shorter wavelengths, the computational box might not be sufficiently high  to encompass enough of the region where the radiation originates. This problem becomes particularly acute for the CN and GB data, due to the presence of strong lines formed particularly high.

%We have also computed simulated facular  contrasts as a function of limb angle. They show a flattening or even turnover at a limb angle of about $\mu=0.3$. The method applied in this work, i.e. calculating contrasts from 3-D MHD simulations using ATLAS9, leads to a potentially more realistic limb behaviour of facular contrasts than previous 1-D modeling.  

The approach presented in this study should allow us to calculate  the facular and network contrast not solely as a function of limb angle, but also as a function  of magnetic flux. For this purpose, we will calculate magnetograms from \textit{MURaM} simulations to identify the previously determined emergent intensities with corresponding magnetic fluxes. Depending on further tests,  these facular and network intensities  could then be incorporated into irradiance reconstruction models (such as SATIRE; see e.g. Krivova et al.~2003) which would enable us to help improve these models, as well as identify the cause of irradiance variations even more reliably.
Of particular interest will also be a detailed comparison of the output of MHD simulations with the high-resolution, low-straylight Sunrise observations \citep{bartholetal2010, solankietal2010}, especially those from the SUFI instrument \citep{gandorferetal2010}, which will go beyond just the comparison of rms values at solar disk centre as in \cite{hirzbergeretal2010}.

%SATIRE (Spectral And Total  Irradiance REconstructions; see e.g. Krivova et al.~2003), which would enable us to remove the single free parameter in the SATIRE model and help improve irradiance reconstructions, as well as identify the cause of irradiance variations even more reliably.

%%%%%%%%%%%%%%%%%%%%%%%%%%%%%%%%%%%%%%%%%%%%%%%%%%%%%%%%%%%%%%%%%
% Acknowledgements
%%%%%%%%%%%%%%%%%%%%%%%%%%%%%%%%%%%%%%%%%%%%%%%%%%%%%%%%%%%%%%%%%

\begin{acknowledgements}
NA acknowledges the SNSF (Swiss National Science Foundation) grant PBEZP2-124350. NA and YCU would also like to acknowledge the support through the NERC SolCli consortium grant (UK) and through ISSI (Switzerland). This work was partially supported by WCU grant No. R31-10016 of the Korean Ministry of Education, Science and Technology. The authors would like to acknowledge Hinode, the Japanese mission, developed and launched by ISAS/JAXA, with NAOJ as domestic partner and NASA and STFC (UK) as international partners. It is operated by these agencies in co-operation with ESA and NSC (Norway).

\end{acknowledgements}

%%%%%%%%%%%%%%%%%%%%%%%%%%%%%%%%%%%%%%%%%%%%%%%%%%%%%%%%%%%%%%%%%
% Bibliography
%%%%%%%%%%%%%%%%%%%%%%%%%%%%%%%%%%%%%%%%%%%%%%%%%%%%%%%%%%%%%%%%%

\bibliographystyle{aa}
\bibliography{journals,nafram}

\begin{thebibliography}{29}
\expandafter\ifx\csname natexlab\endcsname\relax\def\natexlab#1{#1}\fi

\bibitem[{{Barthol} {et~al.}(2010){Barthol}, {Gandorfer}, {Solanki},
  {Sch{\"u}ssler}, {Chares}, {Curdt}, {Deutsch}, {Feller}, {Germerott},
  {Grauf}, {Heerlein}, {Hirzberger}, {Kolleck}, {Meller}, {M{\"u}ller},
  {Riethm{\"u}ller}, {Tomasch}, {Kn{\"o}lker}, {Lites}, {Card}, {Elmore},
  {Fox}, {Lecinski}, {Nelson}, {Summers}, {Watt}, {Mart{\'{\i}}nez Pillet},
  {Bonet}, {Schmidt}, {Berkefeld}, {Title}, {Domingo}, {Gasent Blesa}, {del
  Toro Iniesta}, {L{\'o}pez Jim{\'e}nez}, {{\'A}lvarez-Herrero},
  {Sabau-Graziati}, {Widani}, {Haberler}, {H{\"a}rtel}, {Kampf}, {Levin},
  {P{\'e}rez Grande}, {Sanz-Andr{\'e}s}, \& {Schmidt}}]{bartholetal2010}
{Barthol}, P., {Gandorfer}, A., {Solanki}, S.~K., {et~al.} 2010, Sol. Phys., in
  press

\bibitem[{{Danilovic} {et~al.}(2008){Danilovic}, {Gandorfer}, {Lagg},
  {Solanki}, {V\"ogler}, {Katsukawa}, \& {Tsuneta}}]{danilovicetal2008}
{Danilovic}, S., {Gandorfer}, A., {Lagg}, A., {et~al.} 2008, {\aap}, 484, L17

\bibitem[{{Foukal} \& {Lean}(1988)}]{foukallean1988}
{Foukal}, P. \& {Lean}, J. 1988, ApJ, 328, 347

\bibitem[{{Gandorfer} {et~al.}(2010){Gandorfer}, {Grauf}, {Barthol},
  {Riethmueller}, {Solanki}, {Chares}, {Deutsch}, {Ebert}, {Feller},
  {Germerott}, {Heerlein}, {Heinrichs}, {Hirche}, {Hirzberger}, {Kolleck},
  {Meller}, {Mueller}, {Schaefer}, {Tomasch}, {Knoelker}, {Martinez Pillet},
  {Bonet}, {Schmidt}, {Berkefeld}, {Feger}, {Heidecke}, {Soltau},
  {Tischenberg}, {Fischer}, {Title}, {Anwand}, \&
  {Schmidt}}]{gandorferetal2010}
{Gandorfer}, A., {Grauf}, B., {Barthol}, P., {et~al.} 2010, Sol. Phys., in
  press

\bibitem[{{Haigh}(2007)}]{haigh2007}
{Haigh}, J.~D. 2007, Living Rev.~Solar Physics, 4, 2,
  http://www.livingreviews.org/lrsp-2007-2

\bibitem[{{Hirzberger} {et~al.}(2010){Hirzberger}, {Feller}, {Riethm{\"u}ller},
  {Sch{\"u}ssler}, {Borrero}, {Afram}, {Unruh}, {Berdyugina}, {Gandorfer},
  {Solanki}, {Barthol}, {Bonet}, {Mart{\'{\i}}nez Pillet}, {Berkefeld},
  {Kn{\"o}lker}, {Schmidt}, \& {Title}}]{hirzbergeretal2010}
{Hirzberger}, J., {Feller}, A., {Riethm{\"u}ller}, T.~L., {et~al.} 2010, ApJ,
  Lett., 723, L154

\bibitem[{{Ichimoto} {et~al.}(2008){Ichimoto}, {Katsukawa}, {Tarbell}, {Shine},
  {Hoffmann}, {Berger}, {Cruz}, {Suematsu}, {Tsuneta}, {Shimizu}, \&
  {Lites}}]{ichimotoetal2008}
{Ichimoto}, K., {Katsukawa}, Y., {Tarbell}, T., {et~al.} 2008, in ASP Conf.
  Ser., Vol. 397, First Results From Hinode, ed. S.~A. {Matthews}, J.~M.
  {Davis}, \& L.~K. {Harra}, 51

\bibitem[{{Kosugi} {et~al.}(2007){Kosugi}, {Matsuzaki}, {Sakao}, {Shimizu},
  {Sone}, {Tachikawa}, {Hashimoto}, {Minesugi}, {Ohnishi}, {Yamada}, {Tsuneta},
  {Hara}, {Ichimoto}, {Suematsu}, {Shimojo}, {Watanabe}, {Shimada}, {Davis},
  {Hill}, {Owens}, {Title}, {Culhane}, {Harra}, {Doschek}, \&
  {Golub}}]{kosugietal2007}
{Kosugi}, T., {Matsuzaki}, K., {Sakao}, T., {et~al.} 2007, Sol. Phys., 243, 3

\bibitem[{{Krivova} {et~al.}(2003){Krivova}, {Solanki}, {Fligge}, \&
  {Unruh}}]{krivovaetal2003}
{Krivova}, N.~A., {Solanki}, S.~K., {Fligge}, M., \& {Unruh}, Y.~C. 2003,
  {\aap}, 399, L1

\bibitem[{{Kurucz}(1993)}]{kurucz1993}
{Kurucz}, R.~L. 1993 (CDROM No. 13)

\bibitem[{{Marsh} \& {Svensmark}(2000)}]{marschsvensmark2000}
{Marsh}, N.~D. \& {Svensmark}, H. 2000, Phys. Rev. Lett., 85, 5004

\bibitem[{{Mathew} {et~al.}(2009){Mathew}, {Zakharov}, \&
  {Solanki}}]{mathewetal2009}
{Mathew}, S.~K., {Zakharov}, V., \& {Solanki}, S.~K. 2009, {\aap}, 501, L19

\bibitem[{{Ortiz} {et~al.}(2002){Ortiz}, {Solanki}, {Domingo}, {Fligge}, \&
  {Sanahuja}}]{ortizetal2002}
{Ortiz}, A., {Solanki}, S.~K., {Domingo}, V., {Fligge}, M., \& {Sanahuja}, B.
  2002, {\aap}, 388, 1036

\bibitem[{{Potgieter}(1998)}]{potgieter1998}
{Potgieter}, M.~S. 1998, Space Sci. Rev., 83, 147

\bibitem[{{Press} {et~al.}(1992){Press}, {Teukolsky}, {Vetterling}, \&
  {Flannery}}]{pressetal}
{Press}, W.~H., {Teukolsky}, S.~A., {Vetterling}, W.~T., \& {Flannery}, B.~P.
  1992, ``Numerical Recipes in C", 2nd edn. (New York: Cambridge University
  Press)

\bibitem[{{Shimizu} {et~al.}(2008){Shimizu}, {Nagata}, {Tsuneta}, {Tarbell},
  {Edwards}, {Shine}, {Hoffmann}, {Thomas}, {Sour}, {Rehse}, {Ito},
  {Kashiwagi}, {Tabata}, {Kodeki}, {Nagase}, {Matsuzaki}, {Kobayashi},
  {Ichimoto}, \& {Suematsu}}]{shimizuetal2008}
{Shimizu}, T., {Nagata}, S., {Tsuneta}, S., {et~al.} 2008, Sol. Phys., 249, 221

\bibitem[{{Simpson}(1998)}]{simpson1998}
{Simpson}, J.~A. 1998, Space Sci. Rev., 83, 7

\bibitem[{{Solanki} {et~al.}(2010){Solanki}, {Barthol}, {Danilovic}, {Feller},
  {Gandorfer}, {Hirzberger}, {Riethm{\"u}ller}, {Sch{\"u}ssler}, {Bonet},
  {Mart{\'{\i}}nez Pillet}, {del Toro Iniesta}, {Domingo}, {Palacios},
  {Kn{\"o}lker}, {Bello Gonz{\'a}lez}, {Berkefeld}, {Franz}, {Schmidt}, \&
  {Title}}]{solankietal2010}
{Solanki}, S.~K., {Barthol}, P., {Danilovic}, S., {et~al.} 2010, ApJ, Lett., in
  press

\bibitem[{{Solanki} \& {Fligge}(2002)}]{solankifligge2002}
{Solanki}, S.~K. \& {Fligge}, M. 2002, Advances in Space Research, 29, 1933

\bibitem[{{Suematsu} {et~al.}(2008){Suematsu}, {Tsuneta}, {Ichimoto},
  {Shimizu}, {Otsubo}, {Katsukawa}, {Nakagiri}, {Noguchi}, {Tamura}, {Kato},
  {Hara}, {Kubo}, {Mikami}, {Saito}, {Matsushita}, {Kawaguchi}, {Nakaoji},
  {Nagae}, {Shimada}, {Takeyama}, \& {Yamamuro}}]{suematsuetal2008}
{Suematsu}, Y., {Tsuneta}, S., {Ichimoto}, K., {et~al.} 2008, Sol. Phys., 249,
  197

\bibitem[{{Topka} {et~al.}(1997){Topka}, {Tarbell}, \& {Title}}]{topkaetal1997}
{Topka}, K.~P., {Tarbell}, T.~D., \& {Title}, A.~M. 1997, ApJ, 484, 479

\bibitem[{{Tsuneta} {et~al.}(2008){Tsuneta}, {Ichimoto}, {Katsukawa}, {Nagata},
  {Otsubo}, {Shimizu}, {Suematsu}, {Nakagiri}, {Noguchi}, {Tarbell}, {Title},
  {Shine}, {Rosenberg}, {Hoffmann}, {Jurcevich}, {Kushner}, {Levay}, {Lites},
  {Elmore}, {Matsushita}, {Kawaguchi}, {Saito}, {Mikami}, {Hill}, \&
  {Owens}}]{tsunetaetal2008}
{Tsuneta}, S., {Ichimoto}, K., {Katsukawa}, Y., {et~al.} 2008, Sol. Phys., 249,
  167

\bibitem[{{Unruh} {et~al.}(1999){Unruh}, {Solanki}, \&
  {Fligge}}]{unruhetal1999}
{Unruh}, Y., {Solanki}, S.~K., \& {Fligge}, M. 1999, {\aap}, 345, 635

\bibitem[{{V\"ogler}(2005)}]{voegleretal2005ii}
{V\"ogler}, A. 2005, Mem. Soc. Astron. Ital., 76, 842

\bibitem[{{V\"ogler} {et~al.}(2004){V\"ogler}, {Bruls}, \&
  {Sch\"ussler}}]{voegleretal2004}
{V\"ogler}, A., {Bruls}, J. H. M.~J., \& {Sch\"ussler}, M. 2004, {\aap}, 421,
  741

\bibitem[{{V\"ogler} {et~al.}(2005){V\"ogler}, {Shelyag}, {Sch\"ussler},
  {Cattaneo}, {Emonet}, \& {Linde}}]{voegleretal2005}
{V\"ogler}, A., {Shelyag}, S., {Sch\"ussler}, M., {et~al.} 2005, {\aap}, 429,
  335

\bibitem[{{Wedemeyer-B\"ohm} \& {van der Voort}(2009)}]{wedemeyervoort2009}
{Wedemeyer-B\"ohm}, S. \& {van der Voort}, L.~R. 2009, {\aap}, 503, 225

\bibitem[{{Wenzler} {et~al.}(2005){Wenzler}, {Solanki}, \&
  {Krivova}}]{wenzleretal2005}
{Wenzler}, T., {Solanki}, S.~K., \& {Krivova}, N.~A. 2005, {\aap}, 432, 1057

\bibitem[{{Wenzler} {et~al.}(2006){Wenzler}, {Solanki}, {Krivova}, \&
  {Fr\"ohlich}}]{wenzleretal2006}
{Wenzler}, T., {Solanki}, S.~K., {Krivova}, N.~A., \& {Fr\"ohlich}, C. 2006,
  {\aap}, 460, 583

\end{thebibliography}
%\bibliographystyle{aa}
%\bibliography{nafram}
\end{document}